\documentclass[11pt]{article}

\usepackage[
  shownumpages,               % comment to hide page numbers
  bgcolor={255,253,246},      % box bg color: RGB or xcolor name (e.g. gray, blue, etc.)
  braincolor={157,225,252},    % accent color: RGB or xcolor name (e.g. gray, blue, etc.)
  linkcolor=teal,
  citecolor={144,199,225},  % defaults to braincolor
  urlcolor=blue,            % defaults to blue
  citingstyle=authoryear,     % authoryear: [Ivanov et al., 2023] | numbers: [1] | super: ¹
  bibliostyle=plainnat,       % plainnat, abbrvnat, unsrtnat, ...
  bibfile=references          % .bib file name (without .bib)
]{brainlab}

\usepackage{mathtools}      % \coloneqq and other math helpers (loads amsmath extensions)

\usepackage{multirow}
\usepackage{wrapfig}   % для обтекания
\usepackage{nicefrac}

\setbrainmeta{
  title={The Kuramoto Neural Operator: Learning to Solve PDEs via Coupled Oscillator Dynamics},
  authors={
    Petr Badolia\textsuperscript{1}, Leonid Obukhov\textsuperscript{1}, Dmitry Bylinkin\textsuperscript{1}, Aleksandr Beznosikov\textsuperscript{1,2}
  },
  affiliations={
    \textsuperscript{1}Basic Research of Artificial Intelligence Laboratory (BRAIn Lab) \\
    \textsuperscript{2}Innopolis University
  },
  abstract={
    Operator learning is a rapidly advancing area of computational science. It is particularly well suited to problems where a partial differential equation (PDE) must be solved repeatedly under varying physical configurations. Most existing architectures represent the solution operator in a fixed basis. While this assumption is well aligned with global structures, it is less suitable for phenomena governed by local interactions in physical space. We explore an alternative perspective motivated by the observation that the continuum limit of coupled oscillator systems can describe a broad class of PDEs. Building on this idea, we introduce the Kuramoto Neural Operator (KNO), which represents the solution through the evolution of a latent field of interacting oscillators. Across a diverse collection of PDE benchmarks, KNO achieves strong predictive performance, with improvements over competing approaches. Our experimental evaluation also includes an extensive ablation study that quantifies the contribution of each architectural component incorporated into KNO. Furthermore, we show that the model's prediction error is closely linked to the collective dynamics of the latent oscillators. It varies systematically with their degree of synchronization, providing insights into the underlying mechanisms.
  },
}

\begin{document}

\begin{mainpart}

\section{Introduction}

A large class of problems in modern physics and engineering can be
formulated as solving partial differential equations (PDEs) that govern
the evolution of fields distributed over space and time. Numerical algorithms, including finite-difference
\cite{leveque2007finite}, finite-element \cite{ciarlet2002finite},
finite-volume \cite{moukalled2015finite}, and spectral methods
\cite{canuto2006spectral}, remain the standard tools for accurate
simulation of such systems. However, in many practical settings such as mesh optimization, inverse modeling, and real-time forecasting, the same PDE must be solved repeatedly under varying configurations \cite{ashton2024drivaerml}. Running a solver from scratch for each instance quickly becomes a major computational bottleneck \cite{kovachki2023neural}. Neural operators \cite{kovachki2023neural} are deep learning architectures that address this issue. They learn a mapping between function spaces directly from data. A trained operator approximates the solution for a new PDE with a single forward pass, at a fraction of the cost of running a classical solver \cite{li2023geometry}.

Among neural operator architectures, a prominent line of work represents the target solution operator in a fixed, analytically motivated basis  \cite{li2020fourier, cao2024laplace, lu2021learning}. Such operators impose a global coupling across all spatial locations. While this approach represents smooth, large-scale structure efficiently, it smears out sharp or spatially localized features, such as steep fronts, discontinuities, and moving interfaces \cite{yangriesz}.

Instead of engineering a more expressive fixed basis, one can model the dynamics that generate the solution. In many PDEs, it has fewer effective degrees of freedom than the discretized state space \cite{kochkov2020learning, fresca2021comprehensive}. A low-dimensional trajectory is then enough to reconstruct the full field. Such an approach targets the mechanism that produces the output, not its appearance in a prescribed set of modes. This raises the question: \textit{can operator learning be formulated as learning the latent dynamics itself, rather than a representation in a fixed basis?}

Interacting oscillator systems provide a natural candidate for such latent dynamics. Indeed, wave equations can be interpreted as continuum limits of coupled oscillators \cite{torre201405}. Oscillatory structure also appears in Schr\"odinger-type \cite{sacchetti2020derivation} and transport dynamics \cite{gupta2012one}. More generally, recent studies demonstrate that, in the continuum limit, large ensembles of interacting oscillators converge to nonlinear, nonlocal evolution equations \cite{medvedev2018continuum}. Furthermore, oscillatory structure supplies a built-in measure of synchronization, equivalent to a discrete Dirichlet energy density. Such a quantity can potentially serve as an indicator of prediction reliability that is unavailable in traditional neural operator architectures. These findings suggest that oscillator-based dynamics provide a meaningful inductive bias for neural operator architectures.

Motivated by this perspective, we introduce the \textbf{Kuramoto Neural Operator (KNO)}, which represents the latent state of the neural operator as a field of interacting oscillatory units evolving directly over the physical domain. The resulting multi-layer architecture may be interpreted as a sequence of learned evolution operators acting on this latent dynamical system. By repeatedly alternating local interaction and nonlinear evolution, KNO constructs a latent representation capable of capturing both local transport phenomena and long-range dependencies. Finally, the latent state is decoded into the target function space to approximate the underlying PDE solution operator.

\paragraph{Contributions}
\begin{itemize}
\item We propose a novel perspective on neural operators, in which PDE solution operators are modeled through the hidden evolution of a coupled field of oscillators in physical space. Based on this idea, we introduce the Kuramoto Neural Operator as a concrete instantiation.
 
\item We show that the proposed inductive bias is particularly effective on wave-dominated, oscillatory, and transport-dominated neural operator benchmarks.
 
\item We conduct ablation studies to isolate the contributions of the principal components, as well as a series of experiments aimed at investigating the properties of the proposed architecture.
 
\item We identify a correlation between prediction error and the degree of oscillator synchronization. Because this relationship holds locally, it lets us localize prediction error and flag the regions where it concentrates, without any auxiliary probe or learned decoder.
 
\end{itemize}

\section{Preliminaries}

\paragraph{Neural Operators.} 
 Given input and output function spaces $\mathcal{A}(\mathcal{D}; \mathbb{R}^{d_a})$
and $\mathcal{U}(\mathcal{D}; \mathbb{R}^{d_u})$
 defined over a domain $\mathcal{D} \subset \mathbb{R}^d$, the goal of a neural operator is to approximate
\begin{equation*}
    \mathcal{G}^\dagger: \mathcal{A} \to \mathcal{U}, \quad a \mapsto u,
\end{equation*}
where $u = \mathcal{G}^\dagger(a)$
 denotes the solution corresponding to input
$a$, by a parametric map $\mathcal{G}_\theta \approx \mathcal{G}^\dagger$
 learned from data $\{(a_i, u_i)\}_{i=1}^N$ \cite{kovachki2023neural}.

We employ a finite-dimensional discretization of the domain. Let $\{x_j\}_{j=1}^k \subset \mathcal{D}$
 be a set of
$k$ collocation points. The input function
$a$ is represented by its pointwise evaluations $\mathbf{a} = (a(x_1), \ldots, a(x_k)) \in \mathbb{R}^{k \times d_a}$, and analogously for the output $u$. The learning problem then becomes
\begin{equation*}
    \min_{\theta} \left[ \frac{1}{N} \sum_{i=1}^{N} \bigl\| \mathcal{G}_\theta(\mathbf{a}_i) - \mathbf{u}_i \bigr\|^2 \right].
\end{equation*}

In this formulation, a desirable property is discretization-invariance. One set of parameters should be shared across mesh resolutions, including those not seen during training.

\paragraph{Kuramoto system.}

The classic Kuramoto model \cite{kuramoto1984chemical} describes the dynamics of $M$ coupled phase oscillators:
\begin{equation*}
    \dot{\psi}_i = \omega_i + \frac{K}{M} \sum_{j=1}^{M} \sin(\psi_j - \psi_i),
    \quad i = 1,\ldots,M,
\end{equation*} where $\psi_i \in [0,2\pi)$ is the phase of the $i$-th oscillator,
$\omega_i \in \mathbb{R}$ is its natural frequency drawn from some distribution, and $K \ge 0$ is the global coupling strength.

This formulation naturally extends to arbitrary dimensions $q\in\mathbb{S}^{n-1}$. By replacing the rotational term $\omega_i$ with an arbitrary skew-symmetric matrix $\Omega_i$ and generalizing the uniform all-to-all coupling to an arbitrary coupling kernel $K_{ij}\geq0$, we obtain the following generalized model \cite{chandra2019continuous}:
\begin{equation}\label{eq:main_equation}
        \dot{q}_i = \Omega_i q_i + \mathrm{Proj}_{q_i}\!\left(\sum_j K_{ij} q_j \right),
\end{equation}
where $\mathrm{Proj}_{q_i}$ denotes projection onto the tangent space of the sphere at $q_i$.

\section{Related Work}
\paragraph{Neural Operators.}

One of the earliest neural operator architectures is DeepONet \cite{lu2021learning}, which represents a solution operator as an inner product between embeddings of the input function and the evaluation coordinate. The Fourier Neural Operator \cite{li2020fourier} instead parametrizes the mapping in the frequency domain, and has since been extended in many directions \cite{wen2022u, li2023fourier, liu2025enhancing}, including variants that swap the Fourier transform for other integral transforms \cite{cao2024laplace, yangriesz, lu2026solving}.

Closer to our setting, several methods move beyond a fixed basis. The Latent Field Model \cite{kochkov2020learning} encodes the physical state into a low-dimensional, spatially local latent representation, evolves it in time with a learned derivative network, and decodes back to physical space only at the final step. FNO-DEQ \cite{marwah2023deep} instead represents the steady-state solution implicitly, as the fixed point of a weight-tied FNO obtained by root-finding rather than by unrolling a fixed number of layers. In both cases, the latent evolution is learned and carries no explicit physical structure, while the latent states provide no signal of prediction reliability.

\paragraph{The Kuramoto model.} The Kuramoto model \cite{kuramoto1984chemical} is a canonical model for studying synchronization and collective behavior. Despite its simplicity, the model exhibits a rich repertoire of dynamical behavior, shaped by the coupling structure between oscillators \cite{medvedev2018continuum}. Beyond the classical phase model, a broad literature generalizes Kuramoto dynamics by replacing scalar phases with vector-valued states on higher-dimensional manifolds. The most studied case places oscillators on the unit sphere, with further extensions to hyperbolic spaces, Stiefel manifolds, and Lie groups \cite{chandra2019continuous, jacimovic2024kuramoto, lipton2021kuramoto, lohe2009non, markdahl2021almost}. These results show that Kuramoto-type dynamics can describe systems with internal degrees of freedom on structured state spaces, motivating their use as a model of internal evolution in PDE systems.

Only recently have Kuramoto-inspired dynamical systems begun to attract attention in machine learning. Existing works remain relatively few and have primarily focused on applications such as image processing \cite{miyato2025artificial, song2026kuramoto, xiao2026kuramoto}. These studies suggest that synchronization dynamics may provide a useful computational mechanism beyond their traditional applications.

\begin{figure*}[!ht]
\centering
\includegraphics[width=\linewidth]{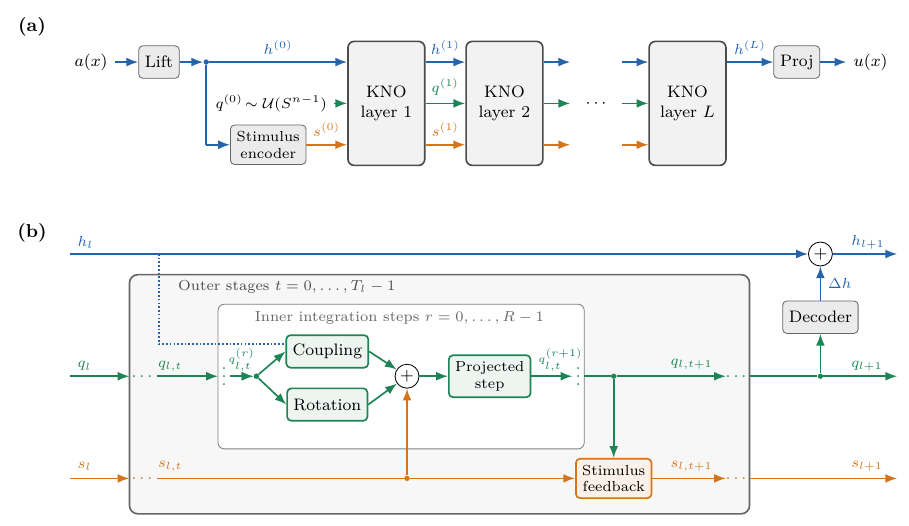}
\caption{KNO architecture.
\textbf{(a)}~The input is lifted pointwise to a feature field $h$
(blue); the oscillators $q$ (green) are initialized at random on the
sphere, and the stimulus $s$ (orange) is initialized from a band-limited
projection of $h$. Next, $L$ KNO layers jointly evolve the triple
$(h,q,s)$, and a pointwise projection produces the output.
\textbf{(b)}~Inside layer $l$, the feature field $h_l$ is resampled to
the canonical grid and conditions the dynamics. Each of the $T_l$ stages integrates the projected Kuramoto update~\eqref{eq:kno-step} for
$R$ integration steps and then refreshes the stimulus through the readout
$g_{l,t}$ in~\eqref{eq:kno-stim}. The final oscillator state is
decoded and added to the feature field as a residual correction
according to~\eqref{eq:kno-dec}.}
\label{fig:kno}
\end{figure*}

\section{Methodology}\label{sec:methodology}
\paragraph{Overview.}
KNO follows the standard neural operator template: a pointwise lifting,
a stack of latent layers, and a pointwise projection. Additionally, several integration stages are also performed within each layer. We index layers by
$l=0,\ldots,L-1$, stages by
$t=0,\ldots,T_l-1$, and integration steps by
$r=0,\ldots,R-1$. Let us further denote a feature grid associated with layer $l$ and a fixed canonical grid as $S_l$ and $S_c$, respectively. Figure~\ref{fig:kno} shows a
schematic overview of the architecture.

Throughout the network, the model maintains three interacting latent
objects: a \emph{feature field} $h_l \colon S_l \to \mathbb{R}^{c}$,
which stores the current representation of the solution; an
\emph{oscillator field} $q_{l,t} \colon S_c \to (\mathbb{S}^{n-1})^M$,
which carries the latent dynamics; and a \emph{stimulus field} $s_{l,t} \colon S_c \to \mathbb{R}^{Mn}$,
through which the feature and oscillator pathways exchange
information. The three fields evolve at different rates: the oscillator field $q$ is updated at every inner step, the stimulus $s$ once per stage, and the feature field $h$ once per layer.

\paragraph{Oscillator state.}
The latent state at each point
$x \in S_c$ consists of $M$ oscillators $\{q_{l,t,m}(x)\}_{m=1}^{M} \subset \mathbb{S}^{n-1}$,
stored jointly as a field with $Mn$ channels. On each forward pass, initial oscillators are sampled independently and uniformly on the sphere by normalizing standard Gaussian vectors.

\paragraph{Stimulus.}
The stimulus is a persistent external conditioning signal. At the
beginning of the network, it is initialized from a band-limited
projection of the lifted input via a convolutional encoder:
\begin{equation*}
\label{eq:kno-stim-init}
s_{0,0}
=
\mathrm{Encoder}
\left(
\mathcal{R}_{S_0 \to S_c}
\left[
\Pi_{\leq k}h_0
\right]
\right),
\end{equation*}
where $\Pi_{\leq k}$ retains the Fourier coefficients within a centered
low-frequency window, $\mathcal{R}_{S_0 \to S_c}$ is the spectral resampling operator. It is implemented by cropping or zero-padding Fourier coefficients.

\paragraph{Canonical latent grid.}
The oscillator and stimulus fields are represented on the fixed
canonical grid $S_c$. At the beginning of layer $l$, the feature field
is transferred from its current grid $S_l$ to $S_c$ using $h_l^c = \mathcal{R}_{S_l \to S_c}[h_l]$. The conditioning field $h_l^c$ is held fixed throughout the oscillator dynamics of layer $l$. Since the evolution of $q_{l,t,m}$ and $s_{l,t}$ is always evaluated on $S_c$, its computational cost is decoupled from the input resolution.

\paragraph{Local coupling.}
KNO replaces the kernel $K_{ij}$ in
\eqref{eq:main_equation} with a learned mapping. At integration step $r$, it concatenates the current oscillator field
$q_{l,t}^{(r)}$ with the conditioning field $h_l^c$ and passes the
result through a convolutional block. The block's output is concatenated
with its original input, and a pointwise MLP then maps this
representation to $Mn$ message channels of $c_{l,t}
(q_{l,t}^{(r)},h_l^c)$.

\paragraph{Rotation and latent dynamics.}
Let $q_{l,t}^{(0)}$ denote the oscillator state at the beginning of stage $t$. At each integration step $r$, the learnable skew-symmetric matrix $\Omega_{l,t,m}$ is applied to corresponding oscillator $m$. The resulting rotation is then combined with the oscillator interaction and the stimulus. Thus, denoting $c_{l,t,m} \equiv c_{l,t}(q_{l,t}^{(r)}, h_l^c)_m$
and dropping arguments for brevity, the force driving inner update $r$
for the $m$-th oscillator is
\begin{equation}
\label{eq:kno-ode}
F_{l,t,m} = c_{l,t,m} + s_{l,t,m} + \Omega_{l,t,m}\,q_{l,t,m}^{(r)}.
\end{equation}
Terms in \eqref{eq:kno-ode} are represented as
fields with $Mn$ channels, corresponding to one $n$-dimensional force
vector per oscillator.

\paragraph{Projected step.}
At integration step $r$, KNO first projects the
oscillator field onto the tangent space at the current state:
\[
v_{l,t,m}^{(r)}
=
\operatorname{Proj}_{q_{l,t,m}^{(r)}}
\left[
F_{l,t,m}
\left(
q_{l,t}^{(r)}, h_l^c, s_{l,t}
\right)
\right].
\]
It then takes an Euler step along this tangent vector and retracts the
result oscillator-wise onto the sphere using
\begin{equation}
\label{eq:kno-step}
q_{l,t,m}^{(r+1)}
=
\mathcal{N}\left(
q_{l,t,m}^{(r)}
+
\gamma_{l,t}v_{l,t,m}^{(r)}
\right),
\end{equation}
where $\mathcal{N}(z) = \nicefrac{z}{\lVert z\rVert}$,
the learned scalar update coefficient $\gamma_{l,t}$ is shared across
all integration steps of the current stage. After $R$ integration steps, the oscillator state at the beginning of the next stage is defined by $q_{l,t+1}=q_{l,t}^{(R)}$.

\paragraph{Stimulus feedback.}

After the $R$ integration steps, the stimulus is
refreshed by an exponential moving average:
\begin{equation}
\label{eq:kno-stim}
s_{l,t+1}
=
\alpha_l s_{l,t}
+
\left(1-\alpha_l\right)
g_{l,t}
\left(
q_{l,t+1}
\right),
\end{equation}
where $\alpha_l=\sigma(a_l)
\in (0,1)$ is a learned layer-specific retention coefficient shared across the
outer stages of layer $l$.

Following the norm-based readout of
\cite{miyato2025artificial}, $g_{l,t}$ applies a learned pointwise
expansion to $q_{l,t+1}$, reshapes the expanded channels into groups,
takes their Euclidean norms, adds a learned bias, and applies a
lightweight convolutional postprocessor. Thus, the stimulus provides
persistent feedback from the current oscillator state while retaining
information injected at earlier stages.

\paragraph{KNO layer.}
At each layer $l$, each stage $t$ consists of $R$
weight-tied updates from \eqref{eq:kno-step}, followed by one
stimulus refresh from \eqref{eq:kno-stim}. The parameters $\left(
c_{l,t},
g_{l,t},
\gamma_{l,t},
\Omega_{l,t}
\right)$ are specific to the stage, while the retention coefficient
$\alpha_l$ is shared across all stages of layer $l$.

After all $T_l$ outer stages, a layer-specific pointwise decoder maps the terminal oscillator state to a feature correction. It is then transferred from the canonical grid $S_c$ to the feature grid $S_{l+1}$ and added
residually:
\begin{equation}
\label{eq:kno-dec}
h_{l+1}
=
\mathcal{R}_{S_l \to S_{l+1}}[h_l]
+
\mathcal{R}_{S_c \to S_{l+1}}
\left[
\mathrm{Decoder}_l
\left(
q_{l,T_l}
\right)
\right].
\end{equation}
The oscillator and stimulus states are carried into the next layer without
reinitialization as $q_{l+1,0}=q_{l,T_l}$ and $s_{l+1,0}=s_{l,T_l}$, respectively.

\section{Positioning Relative to AKOrN}
\label{sec:akorn-positioning}
AKOrN \cite{miyato2025artificial} is the closest architectural precedent for
the latent dynamics of KNO. Both represent hidden states via
sphere-valued oscillators, evolve them through iterative Euler updates with tangent-projected
interactions and oscillator-wise normalization, and read them out through
a rotation-invariant, norm-based map. AKOrN, however, targets discrete
perception, where the oscillator readout feeds directly into
the next layer, whereas KNO adapts the same primitive to operator
learning. The main architectural differences follow from this shift.

Firstly, the coupling in KNO is feature-conditioned. Rather than receiving information only as a fixed additive stimulus, KNO concatenates the oscillator field with the feature field $h_l^c$ before the local convolution. This lets the current solution features modulate the interactions directly. Secondly, KNO keeps the feature field separate from the persistent oscillator and stimulus fields and uses two pathways. Within a layer, the readout feeds back into the dynamics through the stimulus~\eqref{eq:kno-stim}. Across layers, a separate decoder maps the oscillator state into a residual feature update~\eqref{eq:kno-dec}.

There are also numerous smaller differences between KNO and AKOrN. A detailed description of the KNO architecture is provided in Appendix \ref{app:implementation-details}.

\section{Experiments}
\begin{table*}[!ht]
\centering
\caption{Relative $L_2$ errors on benchmark PDEs (lower is better). \textbf{The best} result in each row
is bold and the \underline{second best} is underlined; ties at the displayed precision are
marked jointly.}
\label{tab:main-results}
\small
\resizebox{\linewidth}{!}{%
\begin{tabular}{llcccccccccc}
\hline
Task & Split & KNO (ours) & FNO & RNO & ConvFNO & DeepONet & UNet & CNO & FNO-WT & FNO-DEQ & AKOrN \\
\hline

\textit{Poisson}
& ID
& $\underline{0.0074\pm0.0008}$
& $0.0632 \pm 0.0008$
& $0.0645 \pm 0.0013$
& $0.0524 \pm 0.0177$
& $0.1846 \pm 0.0000$
& $0.0101 \pm 0.0005$
& $\mathbf{0.0034 \pm 0.0001}$
& $0.0602 \pm 0.0014$
& $0.0639 \pm 0.0039$
& $0.0641 \pm 0.0003$ \\
& OOD
& $\underline{0.0189\pm0.0086}$
& $0.0510 \pm 0.0063$
& $0.0486 \pm 0.0013$
& $0.0584 \pm 0.0137$
& $0.1277 \pm 0.0000$
& $0.0412 \pm 0.0202$
& $\mathbf{0.0071 \pm 0.0019}$
& $0.0565 \pm 0.0023$
& $0.0596 \pm 0.0031$
& $0.0632 \pm 0.0066$ \\

\textit{Wave}
& ID
& $\mathbf{0.0032 \pm 0.0012}$
& $0.0218 \pm 0.0005$
& $0.0164 \pm 0.0001$
& $0.0171 \pm 0.0005$
& $0.0470 \pm 0.0021$
& $0.0293 \pm 0.0001$
& $\underline{0.0127 \pm 0.0005}$
& $0.0146 \pm 0.0002$
& $0.0140 \pm 0.0006$
& $0.0186 \pm 0.0007$ \\
& OOD
& $\mathbf{0.0137 \pm 0.0002}$
& $0.0277 \pm 0.0007$
& $0.0215 \pm 0.0001$
& $0.0231 \pm 0.0004$
& $0.0517 \pm 0.0014$
& $0.0339 \pm 0.0007$
& $\underline{0.0198 \pm 0.0005}$
& $0.0208 \pm 0.0002$
& $0.0203 \pm 0.0006$
& $0.0242 \pm 0.0006$ \\

\textit{Allen--Cahn}
& ID
& $\mathbf{0.0013 \pm 0.0001}$
& $0.0043 \pm 0.0001$
& $0.0032 \pm 0.0001$
& $\underline{0.0028 \pm 0.0004}$
& $0.2784 \pm 0.0054$
& $0.0175 \pm 0.0004$
& $0.0101 \pm 0.0004$
& $0.0050 \pm 0.0006$
& $0.0042 \pm 0.0003$
& $0.0283 \pm 0.0016$ \\
& OOD
& $\mathbf{0.0251 \pm 0.0056}$
& $0.0347 \pm 0.0061$
& $\underline{0.0307 \pm 0.0044}$
& $0.0450 \pm 0.0285$
& $0.4917 \pm 0.0175$
& $0.0370 \pm 0.0006$
& $0.0862 \pm 0.0256$
& $0.0325 \pm 0.0046$
& $0.0349 \pm 0.0035$
& $0.1454 \pm 0.0190$ \\

\textit{Cont.\ translation}
& ID
& $\mathbf{0.0007\pm0.0000}$
& $0.0031 \pm 0.0001$
& $0.0031 \pm 0.0001$
& $0.0022 \pm 0.0000$
& $0.0155 \pm 0.0002$
& $0.0026 \pm 0.0004$
& $\underline{0.0017 \pm 0.0001}$
& $0.0024 \pm 0.0001$
& $0.0024 \pm 0.0001$
& $0.0064 \pm 0.0004$ \\
& OOD
& $\mathbf{0.0042\pm0.0004}$
& $0.0159 \pm 0.0129$
& $0.0270 \pm 0.0056$
& $0.0068 \pm 0.0008$
& $0.9724 \pm 0.0393$
& $0.0361 \pm 0.0119$
& $0.0061 \pm 0.0005$
& $\underline{0.0045 \pm 0.0009}$
& $\underline{0.0045 \pm 0.0008}$
& $0.0083 \pm 0.0006$ \\

\textit{Disc.\ translation}
& ID
& $\mathbf{0.0120 \pm 0.0006}$
& $0.0169 \pm 0.0004$
& $0.0173 \pm 0.0004$
& $0.0160 \pm 0.0004$
& $0.1184 \pm 0.0017$
& $0.0207 \pm 0.0002$
& $0.0143 \pm 0.0002$
& $0.0158 \pm 0.0003$
& $0.0157 \pm 0.0003$
& $\underline{0.0140 \pm 0.0001}$\\
& OOD
& $\mathbf{0.0123 \pm 0.0008}$ 
& $0.0208 \pm 0.0005$
& $0.0686 \pm 0.0054$
& $0.0191 \pm 0.0023$
& $0.9181 \pm 0.0280$
& $0.1144 \pm 0.0306$
& $0.0178 \pm 0.0009$
& $0.0174 \pm 0.0012$
& $0.0180 \pm 0.0007$
& $\underline{0.0144 \pm 0.0002}$\\

\textit{Darcy}
& ID
& $\mathbf{0.0141 \pm 0.0013}$
& $0.0389 \pm 0.0011$
& $0.0336 \pm 0.0005$
& $0.0490 \pm 0.0070$
& $0.2810 \pm 0.0135$
& $0.0394 \pm 0.0030$
& $\underline{0.0182 \pm 0.0005}$
& $0.0263 \pm 0.0009$
& $0.0300 \pm 0.0018$
& $0.0374 \pm 0.0052$ \\
& OOD
& $\mathbf{0.0201 \pm 0.0010}$
& $0.0466 \pm 0.0008$
& $0.0375 \pm 0.0004$
& $0.0590 \pm 0.0034$
& $0.1929 \pm 0.0056$
& $0.0536 \pm 0.0059$
& $\underline{0.0241 \pm 0.0005}$
& $0.0330 \pm 0.0001$
& $0.0366 \pm 0.0020$
& $0.0428 \pm 0.0047$ \\

\textit{Navier--Stokes}
& ID
& $\mathbf{0.0950 \pm 0.0023}$
& $0.1181 \pm 0.0025$
& $0.1505 \pm 0.0019$
& $0.1219 \pm 0.0071$
& $0.2674 \pm 0.0040$
& $0.1587 \pm 0.0015$
& $\underline{0.1121 \pm 0.0019}$
& $0.1169 \pm 0.0014$
& $0.1432 \pm 0.0009$
& $0.1923 \pm 0.0051$ \\
& OOD
& $0.3899 \pm 0.1141$
& $\mathbf{0.1867 \pm 0.0015}$
& $0.3534 \pm 0.0086$
& $0.3469 \pm 0.0152$
& $0.6245 \pm 0.0339$
& $0.7759 \pm 0.0431$
& $0.2109 \pm 0.0028$
& $\underline{0.1883 \pm 0.0017}$
& $0.2094 \pm 0.0034$
& $0.6414 \pm 0.1710$ \\

\textit{Airfoil}
& ID
& $0.0112 \pm 0.0001$
& $0.0105 \pm 0.0001$
& $0.0116 \pm 0.0003$
& $0.0119 \pm 0.0004$
& $0.0162 \pm 0.0001$
& $0.0118 \pm 0.0002$
& $\underline{0.0102 \pm 0.0002}$
& $0.0104 \pm 0.0002$
& $\mathbf{0.0101 \pm 0.0001}$
& $0.0192 \pm 0.0005$ \\
& OOD
& $0.0212 \pm 0.0011$
& $0.0188 \pm 0.0005$
& $0.0216 \pm 0.0002$
& $0.0210 \pm 0.0009$
& $0.0286 \pm 0.0002$
& $0.0213 \pm 0.0004$
& $0.0196 \pm 0.0004$
& $\underline{0.0183 \pm 0.0001}$
& $\mathbf{0.0181 \pm 0.0007}$
& $0.0286 \pm 0.0009$ \\
\hline
\end{tabular}%
}
\end{table*}

\subsection{Setup} % TODO: пупупу
\paragraph{Benchmark and evaluation protocol.}
\label{sec:benchmarks}

We evaluate KNO on the Representative PDE Benchmark
\cite{raonic2023convolutional}. The suite contains two-dimensional operator-learning problems spanning elliptic, hyperbolic, parabolic, transport, and
incompressible-flow regimes. Each task is formulated as a
supervised map between full fields sampled on a uniform grid. Unless stated
otherwise, the input and target fields have the same spatial resolution, and
all models are trained and evaluated on identical data splits. Each task
provides out-of-distribution (OOD) and in-distribution (ID) data. The OOD split is generated under a task-specific shift, such as different forcing parameters, coupling strengths, or geometric configurations. Neither ID nor OOD test samples enter architecture, hyperparameter, or checkpoint selection. Detailed task definitions, dataset sizes, and OOD shifts are provided in Appendix \ref{app:task-spec}.

All models are trained for 500 epochs with AdamW using a learning rate
of $10^{-3}$. The primary evaluation metric is the sample-averaged relative $L_2$ error, a standard choice for operator learning \cite{wu2026geopt}:
\begin{equation}
\label{eq:relative-l2}
\mathcal{E}
\left(\theta;\mathcal{F}\right)
=
\frac{1}{|\mathcal{F}|}
\sum_{(a,u)\in\mathcal{F}}
\frac{
\left\|\mathcal{G}_{\theta}(a)-u\right\|_2
}{
\left\|u\right\|_2
},
\end{equation}
where the norms are taken over all spatial locations and output channels.
Reported results are the mean and standard deviation of the
ID or OOD test errors over three seeds.

For each task and each architecture, we select the best model using the same two-stage,
validation-only protocol.

Empirical analysis is conducted on a Linux server utilizing an NVIDIA TESLA A100 with 80 GB of GPU memory. Complete selection budgets, optimization schedule, normalization
procedure, and implementation details are given in Appendix \ref{app:training-protocol}.

\paragraph{Baselines.}
We compare KNO against $9$ baselines: FNO \cite{li2020fourier}, its weight-tied and
deep-equilibrium variants FNO-WT and FNO-DEQ
\cite{marwah2023deep}, RNO
\cite{yangriesz}, ConvFNO \cite{liu2025enhancing},
CNO \cite{raonic2023convolutional},
DeepONet \cite{lu2021learning}, UNet \cite{ronneberger2015u}, and a direct
adaptation of AKOrN to operator learning. Together, these baselines cover the principal
design axes against which we isolate the contribution of the oscillator
dynamics. Architectural and implementation details are provided in Appendix \ref{app:baseline-architectures}.

\subsection{Benchmark Results}

Table~\ref{tab:main-results} establishes KNO as one of the strongest methods
across the suite. It ranks best or second on $7$ and $6$ of the $8$ ID and OOD tasks, respectively. Its advantages are clearest on PDEs with pronounced local spatial structure. On all $5$ of these tasks, KNO ranks first both in and out of distribution.

On \textit{Poisson}, KNO ranks second on both splits. Unlike other PDEs, \textit{Poisson} is a fixed linear map from forcing to solution. Such a stationary structure is well matched by hierarchical convolutions, which explains the superiority of CNO. In \textit{Darcy}, by contrast, the coefficient field changes the operator itself, making the map nonlinear and spatially adaptive, which may explain why KNO stays competitive there despite both problems being elliptic.

\textit{Airfoil} is instead dominated by geometry and boundary effects, for which
uniform canonical-grid latent dynamics may provide a less appropriate
inductive bias. Similarly, despite achieving the best ID result on
\textit{Navier--Stokes}, KNO degrades under the combined resolution and
geometry shift in the OOD split. Overall, KNO provides a broadly effective
inductive bias for operator learning, while its limitations are concentrated
in settings dominated by irregular geometry or substantial structural
distribution shifts.

\subsection{Ablations}
\label{sec:ablations}

We separate the ablations by the architectural question they address. All
tables report ID-test relative $L_2$ at validation-selected checkpoints, mean
$\pm$ standard deviation over three seeds.

\paragraph{Dynamic components and controls.}
Starting from the full architecture,
we remove the local coupling, the rotation or the persistent stimulus. We also consider a no-oscillator control that retains iterative latent refinement while
replacing the oscillator-valued state with an unconstrained feature field.

\begin{wraptable}{r}{0.62\linewidth}
\vspace{-1.0\baselineskip}
\centering
\caption{Dynamic-component ablations. Lower is better.}
\vspace{-0.5\baselineskip}
\label{tab:kno-components}
\small
\resizebox{\linewidth}{!}{%
\begin{tabular}{lcc}
\hline
Variant & \textit{Allen--Cahn} & \textit{Darcy} \\
\hline
KNO
& $\mathbf{0.00134\pm0.00007}$
& $\mathbf{0.0141\pm0.0013}$ \\
KNO w/o coupling
& $0.00217\pm0.00010$
& $0.0639\pm0.0013$ \\
KNO w/o rotation
& $0.00208\pm0.00012$
& $0.0472\pm0.0011$ \\
KNO w/o stimulus
& $0.00184\pm0.00010$
& $0.0156\pm0.0015$ \\
No oscillator
& $0.01888\pm0.02100$
& $0.0818\pm0.0259$ \\
\hline
\end{tabular}%
}
\end{wraptable}

Table~\ref{tab:kno-components} demonstrates that local coupling and rotation are the dominant components, while the stimulus provides a
smaller additional benefit. The no-oscillator control performs substantially
worse and is less stable across seeds, showing that repeated computation alone
does not recover the structured oscillator-inspired dynamics of KNO.

\paragraph{Geometry of the update.}

We compare the complete projected-and-retracted
update with three alternatives: normalization without tangent projection (KNO w/o proj.),
tangent projection without normalization (KNO w/o norm.), and an unconstrained Euclidean
update (KNO w/o both). These controls separate the roles of the individual integration
operations from that of constraining the latent state to the sphere.

\begin{wraptable}{r}{0.62\linewidth}
\vspace{-1.0\baselineskip}
\centering
\caption{Geometric ablations. Lower is better.}
\vspace{-0.5\baselineskip}
\label{tab:kno-geometry}
\small
\resizebox{\linewidth}{!}{%
\begin{tabular}{lcc}
\hline
Variant & \textit{Allen--Cahn} & \textit{Darcy} \\
\hline
KNO
& $0.00134\pm0.00007$
& $\mathbf{0.0141\pm0.0013}$ \\
KNO w/o proj.
& $0.00139\pm0.00002$
& $0.0143\pm0.0007$ \\
KNO w/o norm.
& $0.14795\pm0.00349$
& $18.8719\pm10.8946$ \\
KNO w/o both
& $\mathbf{0.00127\pm0.00015}$
& $0.0187\pm0.0072$ \\
\hline
\end{tabular}%
}
\end{wraptable}

Table~\ref{tab:kno-geometry} reveals that retraction by normalization is the most
essential operation. Removing the tangent projection while retaining
normalization leaves performance almost unchanged on both tasks. By contrast,
applying the tangent projection without retracting the state causes a severe
loss of accuracy, because the discretized trajectory is no longer kept on the
manifold for which the projected vector field is defined.

The unconstrained Euclidean update performs comparably to the spherical update
on \textit{Allen--Cahn}. On \textit{Darcy}, however, the Euclidean parameterization
degrades both the mean accuracy and its stability across seeds. Thus, while the
spherical constraint is not strictly necessary for \textit{Allen--Cahn}, it provides a
more robust inductive bias across tasks.

\subsection{Robustness}
In this subsection, we examine the sensitivity of KNO to the number and dimension of the oscillators. Additional studies of robustness can be found in Appendix~\ref{app:robustness}.

\paragraph{Number of oscillators.}

\begin{wraptable}{r}{0.62\linewidth}
\vspace{-1.0\baselineskip}
\centering
\caption{Oscillator-count sweep on \textit{Allen--Cahn} at fixed $n=4$. Lower is better.}
\vspace{-0.5\baselineskip}
\label{tab:kno-ac-n-osc}
\small
\begin{tabular}{rcrc}
\hline
$M$ & \textit{Allen--Cahn} & $M$ & \textit{Allen--Cahn} \\
\hline
1  & $0.00196 \pm 0.00015$ & 16 & $\mathbf{0.00134 \pm 0.00007}$ \\
2  & $0.00177 \pm 0.00026$ & 32 & $0.00160 \pm 0.00020$ \\
4  & $0.00158 \pm 0.00013$ & 64 & $0.00174 \pm 0.00020$ \\
8  & $0.00143 \pm 0.00001$ &    &                              \\
\hline
\end{tabular}%
\end{wraptable}

Table~\ref{tab:kno-ac-n-osc} varies the number of oscillators $M$ while
fixing their dimension at $n=4$. All other architectural choices are kept
fixed.

Performance improves up to $M=16$ and then deteriorates, showing that the gain
cannot be explained by increasing the latent-channel budget or model capacity
alone.

\paragraph{Dimension of oscillators.}

Table~\ref{tab:kno-ac-osc-dim} varies the oscillator dimension $n$ while
keeping the total oscillator-channel budget $Mn$ fixed. This changes how the latent budget is partitioned between the
number of oscillators and their dimensions.

\begin{wraptable}{r}{0.62\linewidth}
\vspace{-1.0\baselineskip}
\centering
\caption{Oscillator-dimension sweep on \textit{Allen--Cahn} at fixed $Mn=64$. Lower is better.}
\vspace{-0.5\baselineskip}
\label{tab:kno-ac-osc-dim}
\small
\begin{tabular}{rcrc}
\hline
$n$ & \textit{Allen--Cahn} & $n$ & \textit{Allen--Cahn} \\
\hline
1  & $0.05962 \pm 0.00005$ & 16 & $0.00145 \pm 0.00010$ \\
2  & $0.02108 \pm 0.00706$ & 32 & $0.00149 \pm 0.00010$ \\
4  & $\mathbf{0.00134 \pm 0.00007}$ & 64 & $0.00172 \pm 0.00039$ \\
8  & $0.00144 \pm 0.00020$ &    &                              \\
\hline
\end{tabular}%
\end{wraptable}

Very low-dimensional oscillators perform substantially worse. In particular,
$n=1$ corresponds to the degenerate sphere $\mathbb{S}^{0}$ and leaves no
continuous directional degree of freedom for the oscillator dynamics. The
configurations with $4\leq n\leq32$ obtain broadly similar errors. At $n=64$, both the mean error and
its variability increase, suggesting that assigning the entire channel budget
to a single high-dimensional oscillator is not beneficial.

\subsection{Collective Dynamics and Prediction Error}
\label{sec:collective-error}
\paragraph{Analysis protocol.}
In this section, we examine how the predictive behavior of KNO is
reflected in its collective oscillator dynamics. Flattening the nested layer, stage, and integration step structure of Section~\ref{sec:methodology}, we index the oscillator trajectory by a single cumulative update $r = 0, \ldots, \sum_{l=0}^{L - 1} T_l R$, where $r = 0$ is the random initialization and each increment is one weight-tied step. For a spatial neighborhood $\mathcal N(x)$, define
\begin{equation*}
D_{\mathrm{loc}}^{(r)}(x)
=
\frac{1}{2M|\mathcal N(x)|}
\sum_{m=1}^{M}\sum_{y\in\mathcal N(x)}
\left\|q_m^{(r)}(x)-q_m^{(r)}(y)\right\|_2^2.
\end{equation*}
Large values of $D_{\mathrm{loc}}$ indicate
local loss of coherence. Up to a constant determined by the neighborhood
convention, $\nicefrac{D_{\mathrm{loc}}}{(\Delta x)^2}$ is a discrete Dirichlet-energy
density of the latent spherical field, where $\Delta x$ denotes the spacing
of $S_c$.

For each sample $i$ and each integration step $r$, we average the local incoherence map over oscillator states, and compare the resulting $\overline{D}_i^{(r)}$ against the ground truth gradient
magnitude $G_i=\|\nabla u_i\|_2$ and local absolute error
$E_i=\mathbb E[|\widehat u_i-u_i|]$. For $G_i$ and $E_i$, we compute within-sample spatial Spearman rank correlations \cite{kendall1962rank}, denoted as $\rho_G$ and $\rho_E$, respectively. 

We additionally treat the final-depth local incoherence as a ranking score against
a binary top-$10\%$ highest-error label within the valid domain. We use the area under the corresponding precision--recall curve (\textbf{AP}) and the
probability that a positive pixel outranks a random negative one (\textbf{AUROC}). \textbf{AP} is more sensitive to the quality of the top of the ranking than \textbf{AUROC}. For
\textit{Airfoil}, all spatial statistics are restricted to fluid cells after
removing a one-cell margin around the solid airfoil and a four-cell outer
border.

\begin{figure*}[!ht]
\centering
\begin{minipage}[t]{0.325\textwidth}
\centering
\includegraphics[width=\linewidth]{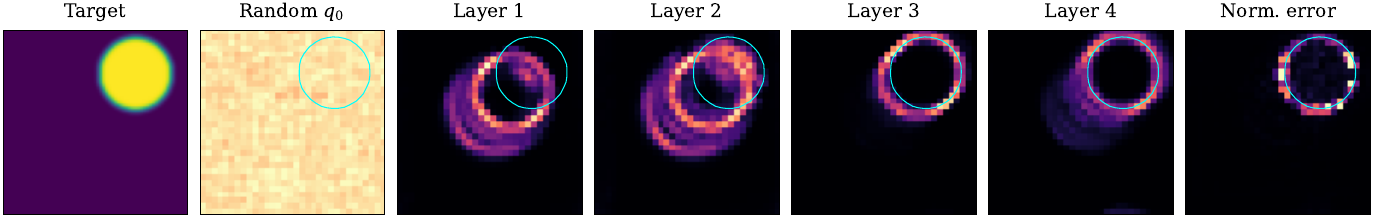}\\[-2pt]
\end{minipage}
\hfill
\begin{minipage}[t]{0.325\textwidth}
\centering
\includegraphics[width=\linewidth]{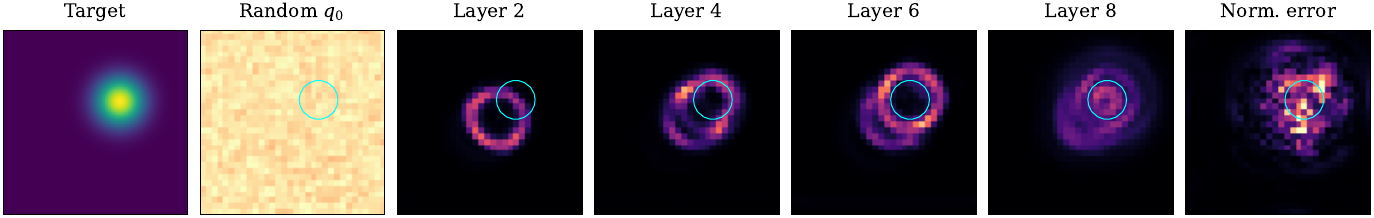}\\[-2pt]
\end{minipage}
\hfill
\begin{minipage}[t]{0.325\textwidth}
\centering
\includegraphics[width=\linewidth]{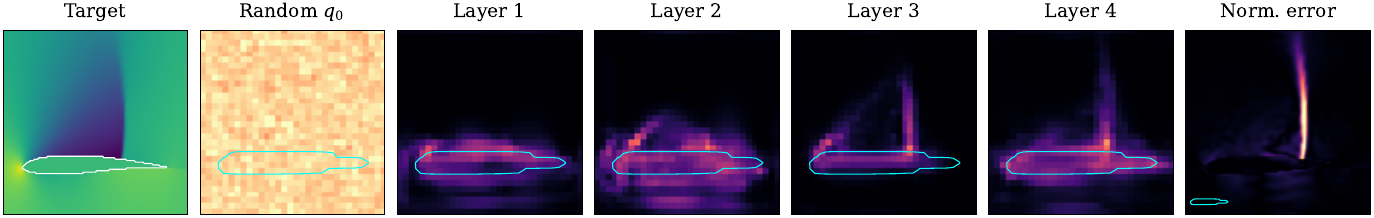}\\[-2pt]
\end{minipage}
\par\vspace{3pt}
\begin{minipage}[t]{0.325\textwidth}
\centering
\includegraphics[width=\linewidth]{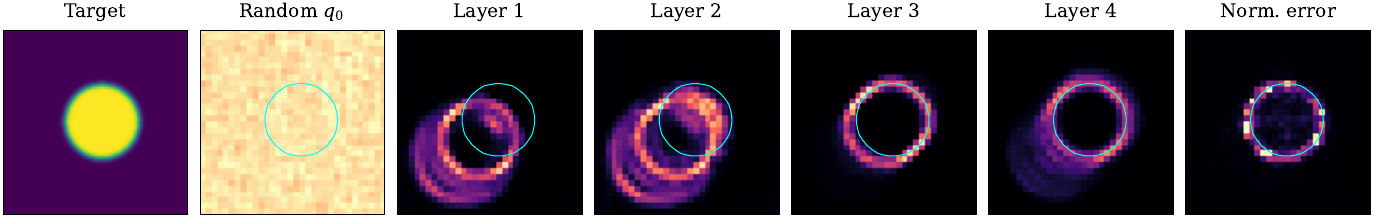}\\[-2pt]
\end{minipage}
\hfill
\begin{minipage}[t]{0.325\textwidth}
\centering
\includegraphics[width=\linewidth]{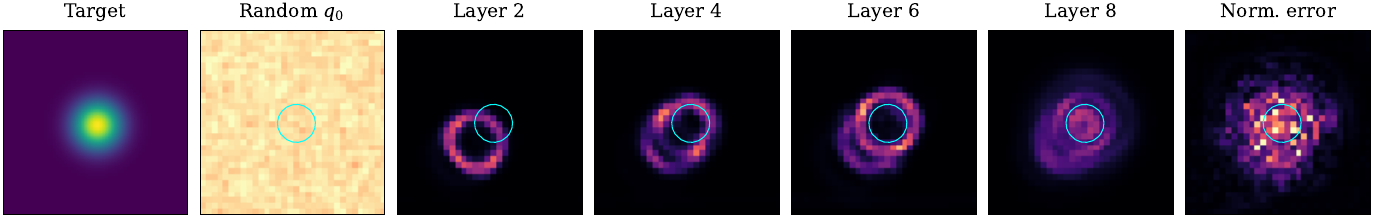}\\[-2pt]
\end{minipage}
\hfill
\begin{minipage}[t]{0.325\textwidth}
\centering
\includegraphics[width=\linewidth]{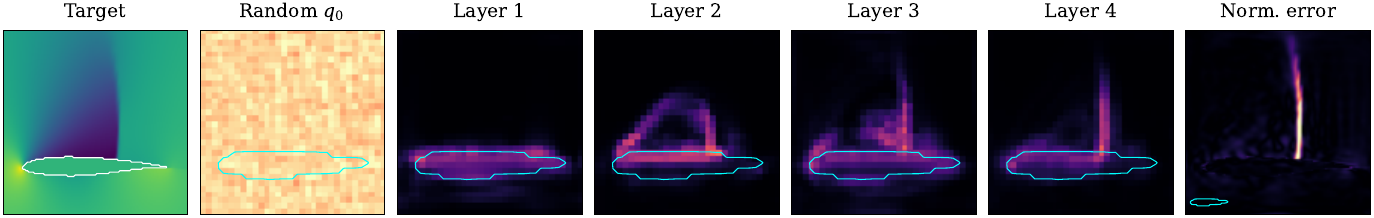}\\[-2pt]
\end{minipage}
\caption{\textbf{Spatial formation of local incoherence across KNO depth.}
Top/bottom rows show OOD/ID examples chosen closest to the median
checkpoint-averaged relative $L_2$ error. Each panel shows the target, the $q_0$-averaged $D_{\mathrm{loc}}$ at initialization and at the indicated layer outputs, and the normalized final error. Translation panels overlay the target
$u{=}0.5$ contour; \textit{Airfoil} panels overlay the airfoil boundary.}
\label{fig:dloc-spatial}
\end{figure*}
\begin{figure*}[!ht]
\centering
\begin{minipage}[!ht]{\textwidth}
\centering
\includegraphics[width=\linewidth]{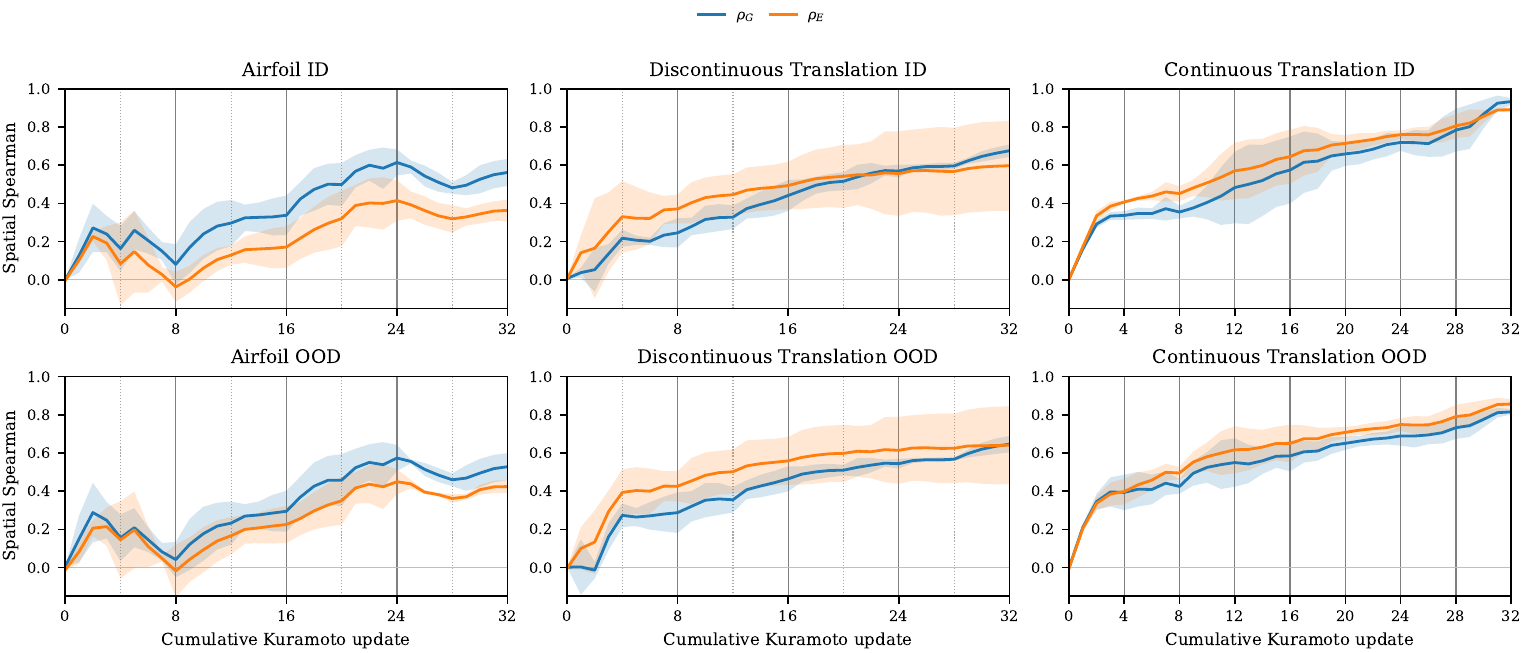}
\end{minipage}
\caption{\textbf{Depthwise formation of geometry and error alignment.}
Curves show median within-sample $\rho_G^{(r)}$ and $\rho_E^{(r)}$; lines and
bands are the mean and standard deviation over three checkpoints.}
\label{fig:dloc-dynamics}
\end{figure*}

\paragraph{Results.}
Initially, local incoherence is diffuse and uninformative. However, through
the KNO layers, it develops into a low-background representation concentrated
near the transported interface on translation tasks and near the airfoil
surface and surrounding high-gradient flow structures on \textit{Airfoil}
(see Figure~\ref{fig:dloc-spatial}). These structures are formed by the learned
dynamics rather than inherited from the random initialization.

\begin{table}[!ht]
\centering
\caption{Final-depth local-incoherence statistics. \textbf{AUROC}/\textbf{AP}: top-$10\%$-error-pixel
ranking (random \textbf{AP} $=0.1$).}
\label{tab:dloc-summary}
\small
\resizebox{\linewidth}{!}{%
\begin{tabular}{llcccc}
\hline
Task & Split & $\rho_G$ & $\rho_E$  & \textbf{AUROC} & \textbf{AP} \\
\hline
\textit{Disc.\ translation} & ID  & $0.676\pm0.027$ & $0.605\pm0.196$ & $0.951\pm0.010$ & $0.782\pm0.022$ \\
 & OOD & $0.646\pm0.035$ & $0.645\pm0.167$ & $0.951\pm0.006$ & $0.782\pm0.018$ \\
\textit{Cont.\ translation} & ID  & $0.933\pm0.015$ & $0.899\pm0.011$ & $0.949\pm0.008$ & $0.630\pm0.038$ \\
 & OOD & $0.815\pm0.011$ & $0.862\pm0.020$ & $0.945\pm0.006$ & $0.590\pm0.028$ \\
\textit{Airfoil} & ID  & $0.562\pm0.071$ & $0.364\pm0.056$ & $0.815\pm0.012$ & $0.356\pm0.041$ \\
 & OOD & $0.528\pm0.071$ & $0.424\pm0.033$ & $0.833\pm0.004$ & $0.378\pm0.026$ \\
\hline
\end{tabular}%
}
\end{table}

At final depth, $D_{\mathrm{loc}}$ is positively aligned with target
geometry and local error on all three tasks, both for ID and OOD splits (Table~\ref{tab:dloc-summary}).
\textit{Continuous Translation} has the strongest domain-wide correlations.
\textit{Discontinuous Translation} has a lower $\rho_G$, reflecting its thin jump
set, but a comparable error relation and the highest \textbf{AP}: its error is
concentrated in a narrow band that is captured by the strongest
$D_{\mathrm{loc}}$ values. \textit{Airfoil} is more difficult because neither target
gradients nor prediction error are confined to a single transported
interface. Nevertheless, $D_{\mathrm{loc}}$ achieves an \textbf{AUROC} $0.815$ on ID and
$0.833$ on OOD. 

Figure~\ref{fig:dloc-dynamics} confirms that these relations are constructed
by the learned dynamics: $\rho_E$ is essentially zero at $r=0$ on every task
and split. On the translation tasks, geometry-controlled error alignment
appears early and then remains stable or grows through depth. On \textit{Airfoil}, the
signal forms later, emerging clearly by the third layer and remaining stable
through the final output.

Thus, the recurrent Kuramoto updates transform an initially random
oscillator field into a task-specific geometric
representation. It acts
as an architecture-native local error-risk signal that retains useful
localization under distribution shift. Additional experiments on collective dynamics are given in Appendix~\ref{app:add-interp}.

\section{Conclusion}
We introduced the Kuramoto Neural Operator, which models a solution operator
through the evolution of a latent field of coupled spherical oscillators.
Across eight PDE benchmarks, KNO ranks first or second on most ID and OOD
tasks, with its strongest results on wave propagation, reaction--diffusion,
translation, and heterogeneous elliptic flow. Ablations show that
local coupling and rotational dynamics provide the largest gains, with a
smaller additional benefit from the persistent stimulus; the spherical
parameterization may not be essential on some tasks, but stabilizes overall performance. The recurrent dynamics turn an initially random oscillator field into
a task-adapted representation of solution geometry, and its local incoherence
localizes prediction errors, remaining informative under
distribution shift without an auxiliary probe or learned uncertainty head.

\section{Limitations}
KNO evolves on a uniform grid. It is therefore less suited to problems
dominated by irregular geometry, which is visible in our experiments on the
\textit{Airfoil} task. At the opposite end of the spectrum, KNO offers
little advantage on simple linear operators, whose solution is a global smoothing of the input and is
already captured well by standard convolutional models.

We have also restricted ourselves to a principled architectural
design and have not explored several more advanced mechanisms, for instance,
attention-based coupling. We
expect that a more thorough investigation of each individual component of KNO could
yield further gains and, in particular, help address the tasks dominated
by irregular geometry.

Finally, the present evaluation is limited to two-dimensional, grid-based
operators from a single benchmark suite. Establishing how KNO behaves on unstructured
domains, high-dimensional problems, and long-horizon temporal rollout remains an open
direction.

\end{mainpart}

\begin{appendixpart}

%\onecolumn
%\appendix
%\setcounter{secnumdepth}{2}

\section{Robustness}
\label{app:robustness}

We complement the oscillator-count and oscillator-dimension studies in
Section~\ref{sec:ablations} with additional analyses of integration depth,
parameterized depth, random initialization, and resolution transfer. Unless
stated otherwise, all tables report ID-test relative $L_2$ errors at
validation-selected checkpoints, mean $\pm$ standard deviation over three
seeds.

\subsection{Number of Integration Steps per Stage} % ???
\label{app:kno-ac-micro-steps}

Table~\ref{tab:kno-ac-micro-steps} varies the number $R$ of weight-tied
integration steps within a stage of KNO. The vector-field
parameters are shared across these steps. Increasing $R$ therefore changes
the integration depth without introducing additional
parameters.

\begin{wraptable}{r}{0.62\linewidth}
\vspace{-1.0\baselineskip}
\centering
\caption{Integration step sweep on \textit{Allen--Cahn}. Lower is better.}
\vspace{-0.5\baselineskip}
\label{tab:kno-ac-micro-steps}
\small
\begin{tabular}{rc@{\qquad}rc}
\hline
$R$ & \textit{Allen--Cahn} & $R$ & \textit{Allen--Cahn} \\
\hline
1  & $0.00172 \pm 0.00019$
& 8  & $0.00132 \pm 0.00009$ \\
2  & $0.00146 \pm 0.00004$
& 12 & $0.00128 \pm 0.00014$ \\
3  & $0.00138 \pm 0.00002$
& 16 & $0.00127 \pm 0.00008$ \\
4  & $0.00134 \pm 0.00007$
& 32 & $0.00121 \pm 0.00004$ \\
6  & $0.00132 \pm 0.00011$
& 64 & $0.00117 \pm 0.00003$ \\
\hline
\end{tabular}%
\end{wraptable}

The error decreases sharply as $R$ grows from $1$ to $3$, falling from $0.00172\pm0.00019$ to $0.00138\pm0.00002$. Beyond this range, the gains diminish substantially. Indeed, the settings $4\leq R\leq32$ produce values that lie within seed-to-seed variability. Thus, even a small number of integration steps is sufficient to reach an accurate latent trajectory.

\subsection{Number of Stages per Layer}
\label{app:kno-ac-cells-per-layer}

Table~\ref{tab:kno-ac-cells-per-layer} varies the number $T$ of independently
parameterized Kuramoto stages in each KNO layer.

\begin{wraptable}{r}{0.32\linewidth}
\vspace{-1.0\baselineskip}
\centering
\caption{Stage count sweep on \textit{Allen--Cahn}. Lower is better.}
\vspace{-0.5\baselineskip}
\label{tab:kno-ac-cells-per-layer}
\small
\begin{tabular}{rc}
\hline
$T$ & \textit{Allen--Cahn} \\
\hline
1 & $0.00148 \pm 0.00007$ \\
2 & $0.00134 \pm 0.00007$ \\
3 & $0.00129 \pm 0.00010$ \\
4 & $0.00128 \pm 0.00005$ \\
\hline
\end{tabular}%
\end{wraptable}

We set $(T_0,T_1,T_2,T_3)=(T,T,T,T)$ and keep the number of integration steps within each stage fixed.
Unlike increasing $R$, increasing $T$ introduces additional vector fields and
therefore increases both the parameter count and the depth of the
parameterized dynamics.

Additional parameterized depth improves accuracy, but the gains saturate
rapidly. The largest reduction occurs between $T=1$ and $T=2$, whereas the
results for $T=3$ and $T=4$ are indistinguishable relative to their
seed-to-seed variability. This suggests that the number of stages within a layer should not be chosen to be too large during tuning, same as the number of integration steps within a stage.

\subsection{Number of Layers}
\label{app:kno-ac-layers}

\begin{wraptable}{r}{0.32\linewidth}
\vspace{-1.0\baselineskip}
\centering
\caption{Layer-count sweep on \textit{Allen--Cahn}. Lower is better.}
\vspace{-0.5\baselineskip}
\label{tab:kno-ac-layers}
\small
\begin{tabular}{rc}
\hline
$L$ & \textit{Allen--Cahn} \\
\hline
1 & $0.00199 \pm 0.00027$ \\
2 & $0.00150 \pm 0.00002$ \\
4 & $0.00134 \pm 0.00007$ \\
6 & $0.00138 \pm 0.00013$ \\
\hline
\end{tabular}
\end{wraptable}

Table~\ref{tab:kno-ac-layers} varies the number $L$ of KNO layers while
keeping the remaining architectural and optimization settings fixed.
Increasing $L$ adds independently parameterized feature updates and extends
the latent oscillator trajectory across the network.

Increasing the depth from $1$ to $2$ layers substantially reduces the error,
and $4$ layers also provide improvement. Further extending the network to $6$
layers yields no additional gain and slightly increases seed-to-seed
variability. Consequently, as with the oscillator dimensionality and count, there exists a trade-off in the number of layers beyond which the optimization becomes increasingly difficult, leading to diminishing performance gains.

\subsection{Sensitivity to Random Initialization}

We evaluate sensitivity to the random oscillator initialization on the
\textit{Allen--Cahn} ID-test set. For the selected seed-0 checkpoint, we run
inference on 64 samples using 16 independent realizations of oscillator field initialization $q_0$. The mean
relative pointwise standard deviation of the prediction is
$1.30\times10^{-4}$, and the mean pairwise relative prediction disagreement
is $1.86\times10^{-4}$. The standard deviation of the dataset-level relative
$L_2$ error across initializations is only $1.25\times10^{-6}$, compared with
a mean error of $1.22\times10^{-3}$. Averaging predictions over the 16
initializations reduces the relative $L_2$ error by only $0.59\%$. Thus,
although $q_0$ is sampled randomly, its effect on the prediction is negligible,
and the trained operator is effectively deterministic at the scale of its
prediction error.

\subsection{Zero-Shot Resolution Transfer}
\label{app:resolution}

\begin{figure}[!ht]
\centering
\begin{minipage}[t]{0.49\textwidth}
    \centering
    \includegraphics[width=\linewidth]{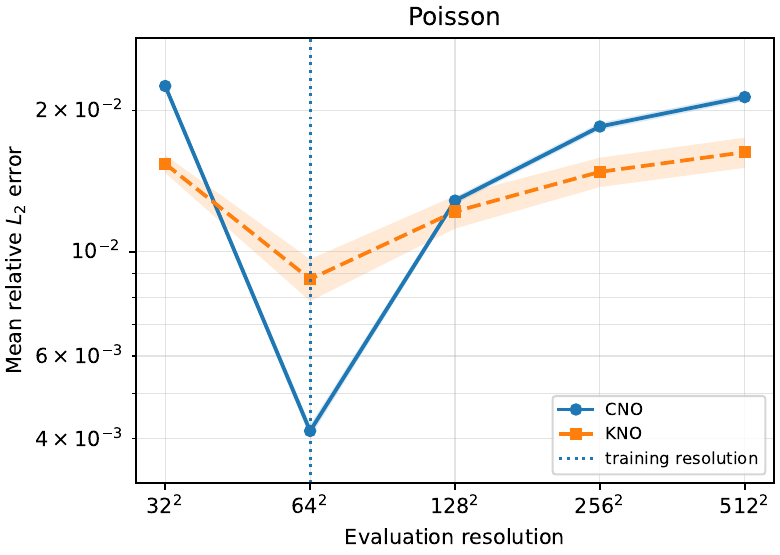}
    \caption{Zero-shot resolution transfer on \textit{Poisson}. Curves show
    mean relative $L_2$ error and shaded regions show one standard deviation. We display KNO and the strongest competitor. The dashed line marks the
    training resolution.}
    \label{fig:resolution-transfer-poisson}
\end{minipage}%
\hfill
\begin{minipage}[t]{0.49\textwidth}
    \centering
    \includegraphics[width=\linewidth]{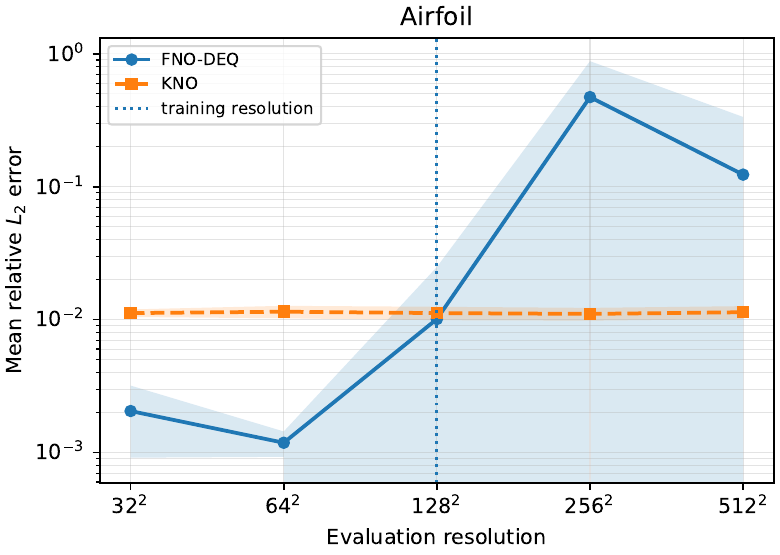}
    \caption{Zero-shot resolution transfer on \textit{Airfoil}. Curves show
    mean relative $L_2$ error and shaded regions show one standard deviation. We display KNO
    and the strongest competitor. The dashed line marks the training resolution.}
    \label{fig:resolution-transfer-airfoil}
\end{minipage}
\end{figure}
    
In this section, we evaluate zero-shot transfer on \textit{Poisson} and \textit{Airfoil},
two benchmarks on which KNO does not lead at the training resolution in
Table~\ref{tab:main-results}. Each
model is evaluated on the same family of solutions at multiple grid sizes
without finetuning.

On \textit{Poisson}, we provide a comparison with CNO, the only architecture that beats KNO at the $64^2$ training grid. KNO
achieves lower error at every evaluated off-grid resolution. Notably, the gap between KNO and CNO widens as the evaluation resolution moves farther away from the training resolution. Thus, although KNO is not the strongest fixed-resolution solver for this linear elliptic operator, its canonical-grid dynamics transfer
more robustly across discretizations.

On \textit{Airfoil}, the KNO curve suffers less than $12\%$ relative change across resolutions. FNO-DEQ attains a lower benchmark error at the native resolution,
but exhibits severe
instability on finer grids. 

Taken together, these results separate fixed-resolution accuracy from
resolution robustness. KNO is substantially
more stable under resolution change on both problems. It retains a stable approximation under substantial variations of grid size.

\section{Resolution Transfer of Local Error Localization}
\label{app:add-interp}

The zero-shot experiments in Appendix~\ref{app:resolution} test whether KNO
retains predictive accuracy away from the training discretization. We now ask whether the local-incoherence signal analyzed at the training
resolution in Section~\ref{sec:collective-error}, including its spatial
formation in Figure~\ref{fig:dloc-spatial} and its depthwise statistics in
Figure~\ref{fig:dloc-dynamics}, also transfers across resolutions. This is a
strictly post-hoc, probe-free analysis. The KNO parameters are frozen, and no
resolution-specific calibration or learned error predictor is introduced.

\subsection{Protocol}

We analyze \textit{Continuous Translation}, \textit{Discontinuous
Translation}, and \textit{Airfoil} at resolutions
$32^2,64^2,128^2,256^2,$ and $512^2$. We
use $64$ samples for each translation task and $32$ samples for
\textit{Airfoil}, evaluated with the same three validation-selected
checkpoints as in the main experiments. Each field is averaged over $4$
independent oscillator initializations $q_0$.

The oscillator dynamics remain on the fixed $32\times32$ canonical grid at
every input resolution. We therefore compute the final-depth local
incoherence $D_{\mathrm{loc}}$ on this grid and transfer the native-grid target
gradient and absolute prediction-error maps to the canonical grid by area
resampling. The translation masks exclude two canonical cells at each outer
boundary. For \textit{Airfoil}, we retain fluid cells only, remove a one-cell
margin around the solid body, and exclude a $4$-cell outer border, exactly as
in Section~\ref{sec:collective-error}.

For each field, we report the within-field Spearman correlations of final
$D_{\mathrm{loc}}$ with target-gradient magnitude and local absolute error,
denoted by $\rho_G$ and $\rho_E$. We also rank the top-$10\%$ highest-error
pixels using $D_{\mathrm{loc}}$ and report \textbf{AUROC} and average precision (\textbf{AP}).
Statistics are summarized by the median over fields within each checkpoint
and then by the mean $\pm$ population standard deviation over the three
checkpoints. Because this experiment uses paired resolution-transfer subsets
rather than the complete ID test splits, its native-resolution rows are not
replacements for Table~\ref{tab:dloc-summary}.

\begin{table}[!ht]
\centering
\caption{Resolution transfer of final-depth local incoherence. \textbf{AUROC}
and \textbf{AP} evaluate the ranking of the top-$10\%$ local-error pixels;
random \textbf{AP} is approximately $0.1$. All entries are mean $\pm$ standard
deviation over three checkpoints.}
\label{tab:resolution-dloc}
\small
\setlength{\tabcolsep}{5.3pt}
\begin{tabular}{llcccc}
\toprule
Task & Resolution & $\rho_G$ & $\rho_E$ & \textbf{AUROC} & \textbf{AP} \\
\midrule
\textit{Cont. translation}
& $32^2$
& $0.922\pm0.020$
& $0.896\pm0.004$
& $0.953\pm0.011$
& $0.619\pm0.058$ \\
& $64^2$ (train)
& $0.932\pm0.015$
& $0.900\pm0.012$
& $0.966\pm0.010$
& $0.747\pm0.072$ \\
& $128^2$
& $0.936\pm0.012$
& $0.919\pm0.013$
& $0.954\pm0.011$
& $0.609\pm0.065$ \\
& $256^2$
& $0.937\pm0.011$
& $0.921\pm0.015$
& $0.954\pm0.011$
& $0.606\pm0.065$ \\
& $512^2$
& $0.937\pm0.011$
& $0.923\pm0.015$
& $0.954\pm0.011$
& $0.608\pm0.070$ \\
\midrule
\textit{Disc. translation}
& $32^2$
& $0.643\pm0.024$
& $0.578\pm0.019$
& $0.942\pm0.006$
& $0.676\pm0.018$ \\
& $64^2$ (train)
& $0.659\pm0.029$
& $0.591\pm0.197$
& $0.951\pm0.009$
& $0.781\pm0.024$ \\
& $128^2$
& $0.693\pm0.024$
& $0.572\pm0.194$
& $0.960\pm0.008$
& $0.803\pm0.021$ \\
& $256^2$
& $0.689\pm0.026$
& $0.575\pm0.191$
& $0.963\pm0.006$
& $0.813\pm0.019$ \\
& $512^2$
& $0.671\pm0.024$
& $0.576\pm0.187$
& $0.963\pm0.006$
& $0.814\pm0.019$ \\
\midrule
\textit{Airfoil}
& $32^2$
& $0.749\pm0.013$
& $0.227\pm0.062$
& $0.839\pm0.046$
& $0.348\pm0.073$ \\
& $64^2$
& $0.728\pm0.028$
& $0.224\pm0.120$
& $0.834\pm0.039$
& $0.399\pm0.128$ \\
& $128^2$ (train)
& $0.700\pm0.054$
& $0.303\pm0.185$
& $0.830\pm0.006$
& $0.359\pm0.074$ \\
& $256^2$
& $0.682\pm0.067$
& $0.319\pm0.199$
& $0.835\pm0.010$
& $0.352\pm0.074$ \\
& $512^2$
& $0.678\pm0.077$
& $0.326\pm0.204$
& $0.840\pm0.008$
& $0.355\pm0.071$ \\
\bottomrule
\end{tabular}
\end{table}

\begin{figure}[!ht]
\centering
\begin{minipage}[t]{0.49\textwidth}
\centering
\includegraphics[width=\linewidth]{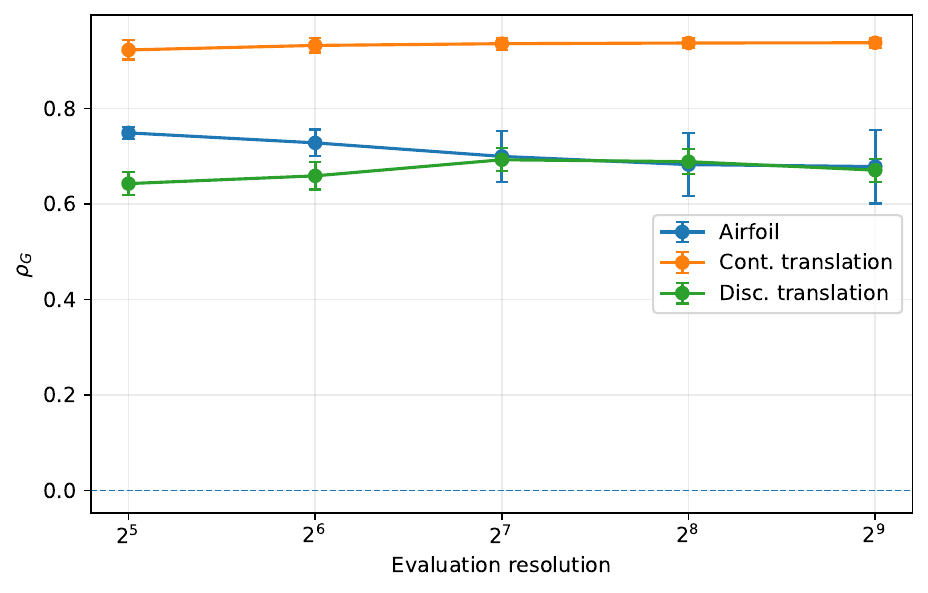}
\end{minipage}
\hfill
\begin{minipage}[t]{0.49\textwidth}
\centering
\includegraphics[width=\linewidth]{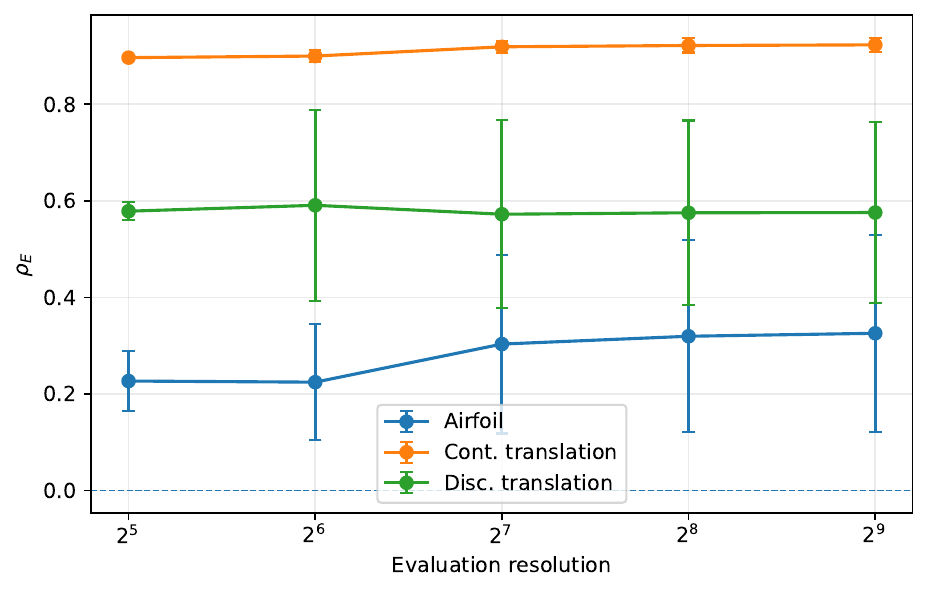}
\end{minipage}

\begin{minipage}[t]{0.49\textwidth}
\centering
\includegraphics[width=\linewidth]{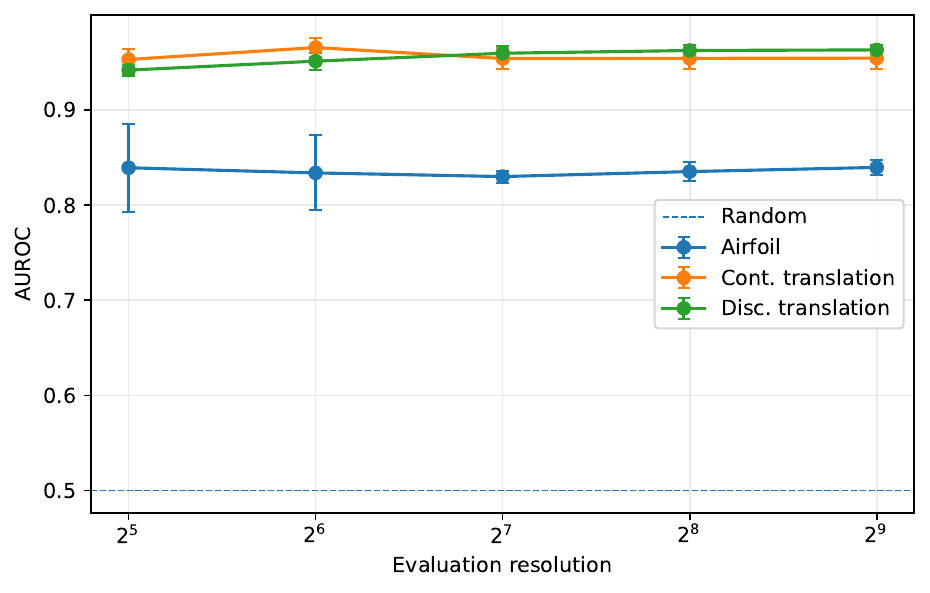}
\end{minipage}
\hfill
\begin{minipage}[t]{0.49\textwidth}
\centering
\includegraphics[width=\linewidth]{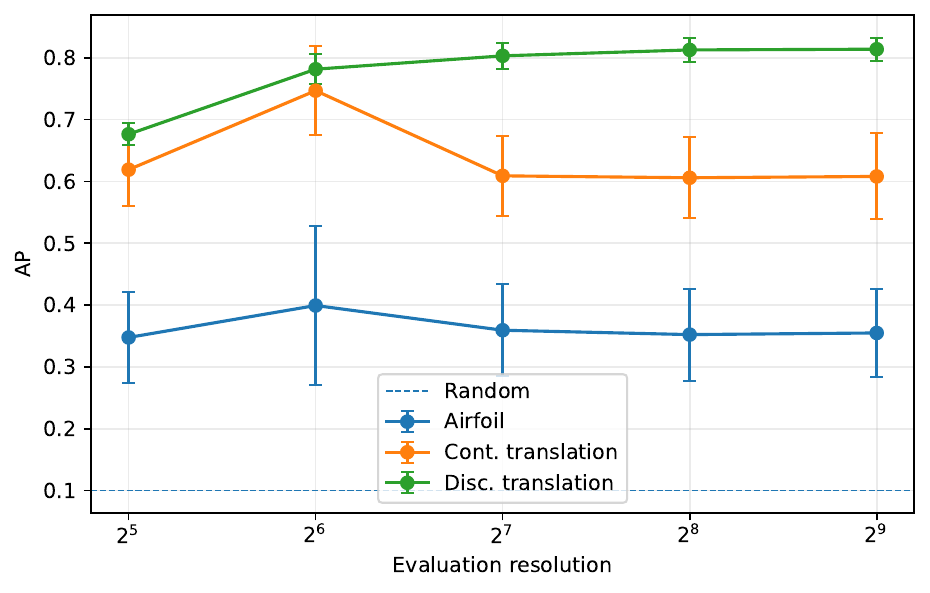}
\end{minipage}
\caption{Resolution dependence of final-depth local-incoherence statistics.
Curves and bands show the mean and standard deviation over three checkpoints.
The local-risk ranking remains informative throughout the evaluated range,
although \textbf{AP} changes in a task-dependent manner.}
\label{fig:resolution-dloc-summary}
\end{figure}

\subsection{Quantitative Results}

Table~\ref{tab:resolution-dloc} and
Figure~\ref{fig:resolution-dloc-summary} summarize the resolution dependence
of the final-depth localization statistics. The local-incoherence ranking
remains informative at every evaluated resolution. Across all task--resolution pairs, \textbf{AUROC} is at least $0.830$.

The clearest example is \textit{Cont. Translation}. From $64^2$ to $512^2$,
$\rho_E$ increases from $0.900$ to $0.923$, while \textbf{AUROC} changes only
from $0.966$ to $0.954$. \textbf{AP} decreases from $0.747$ on the training
grid to approximately $0.61$ off-grid. The most extreme residual pixels are
therefore localized less sharply, despite the stable domain-wide monotone
relation.

\textit{Disc. translation} shows the complementary behavior. \textbf{AUROC}
improves from $0.951$ to $0.963$ and \textbf{AP} from $0.781$ to $0.814$
between $64^2$ and $512^2$. The residual is concentrated around a thin
transported jump set, which is represented more sharply on finer grids. The
mean $\rho_E$ remains near $0.58$ across resolutions, but its large
checkpoint-to-checkpoint standard deviation shows that domain-wide rank
correlation is less stable than the top-error-region metrics on this task.
Accordingly, \textbf{AUROC} and \textbf{AP} provide the cleaner measure of transferred localization for a narrow discontinuity.

For \textit{Airfoil}, the high-error ranking is stable across resolutions:
\textbf{AUROC} stays between $0.830$ and $0.840$, while \textbf{AP} stays near
$0.35$. The error correlation increases slightly from $0.303$ to $0.326$, but
the large checkpoint variability precludes interpreting this difference as a
systematic improvement. The more robust conclusion is that the near-body and
wake-associated risk ranking persists across resolution, whereas a single
domain-wide monotone relation explains the error less consistently than on
the translation tasks.

Figure~\ref{fig:resolution-dloc-samples} shows representative examples of the
target, local absolute error, and final-depth $D_{\mathrm{loc}}$.

\begin{figure}[!ht]
\centering
\begin{minipage}[t]{0.33\textwidth}
\centering
\includegraphics[width=\linewidth]{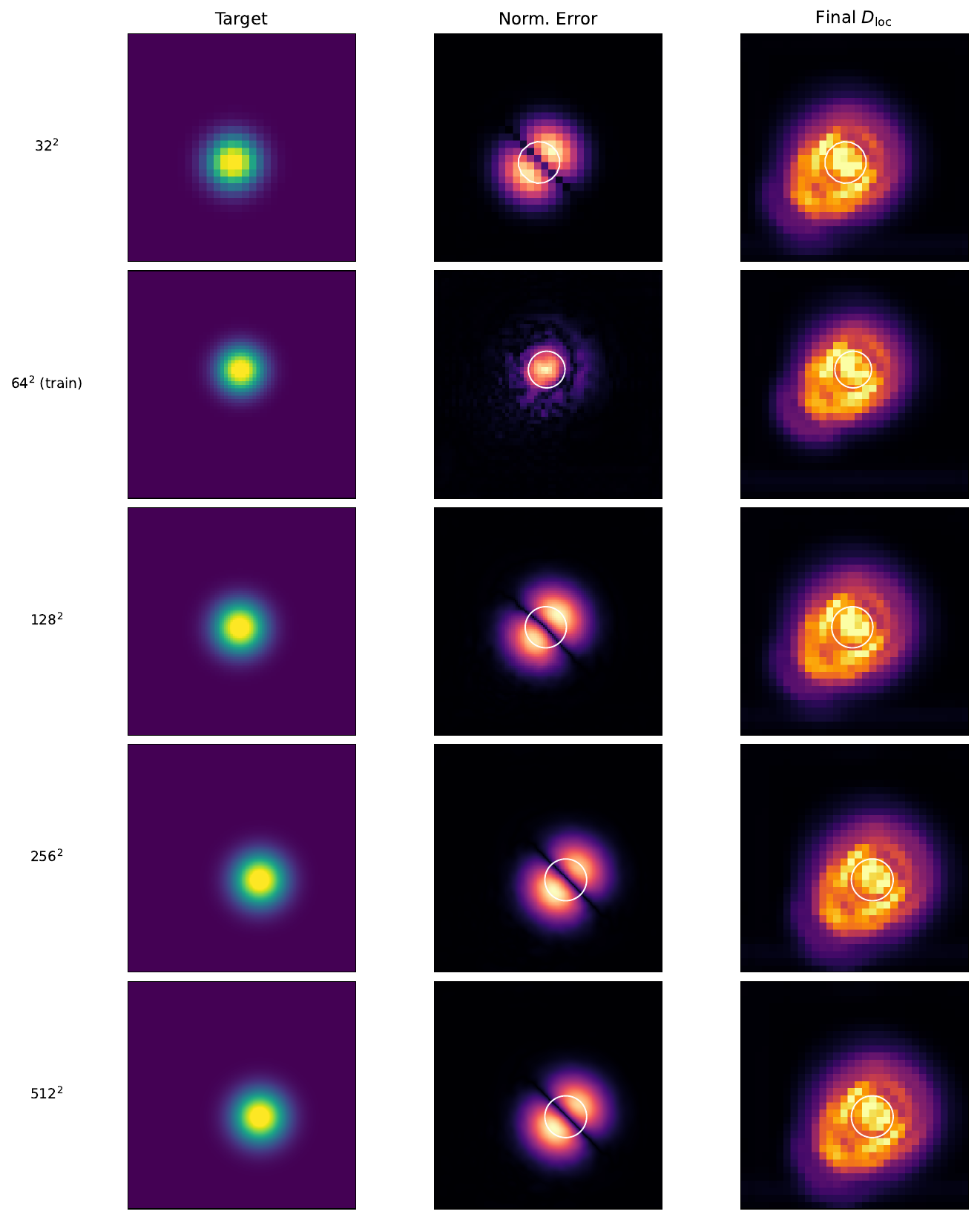}
\end{minipage}%
\hfill
\begin{minipage}[t]{0.33\textwidth}
\centering
\includegraphics[width=\linewidth]{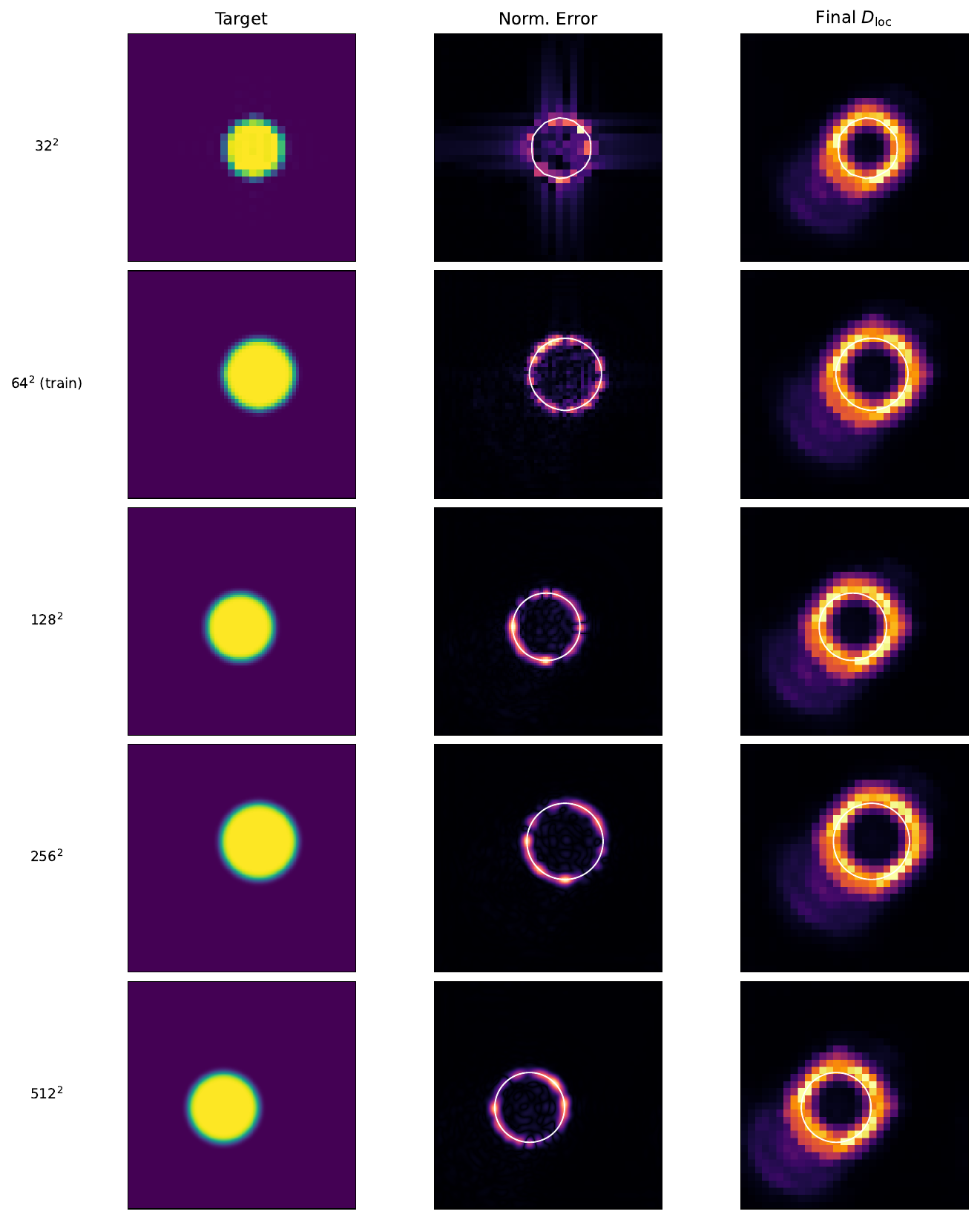}
\end{minipage}%
\hfill
\begin{minipage}[t]{0.33\textwidth}
\centering
\includegraphics[width=\linewidth]{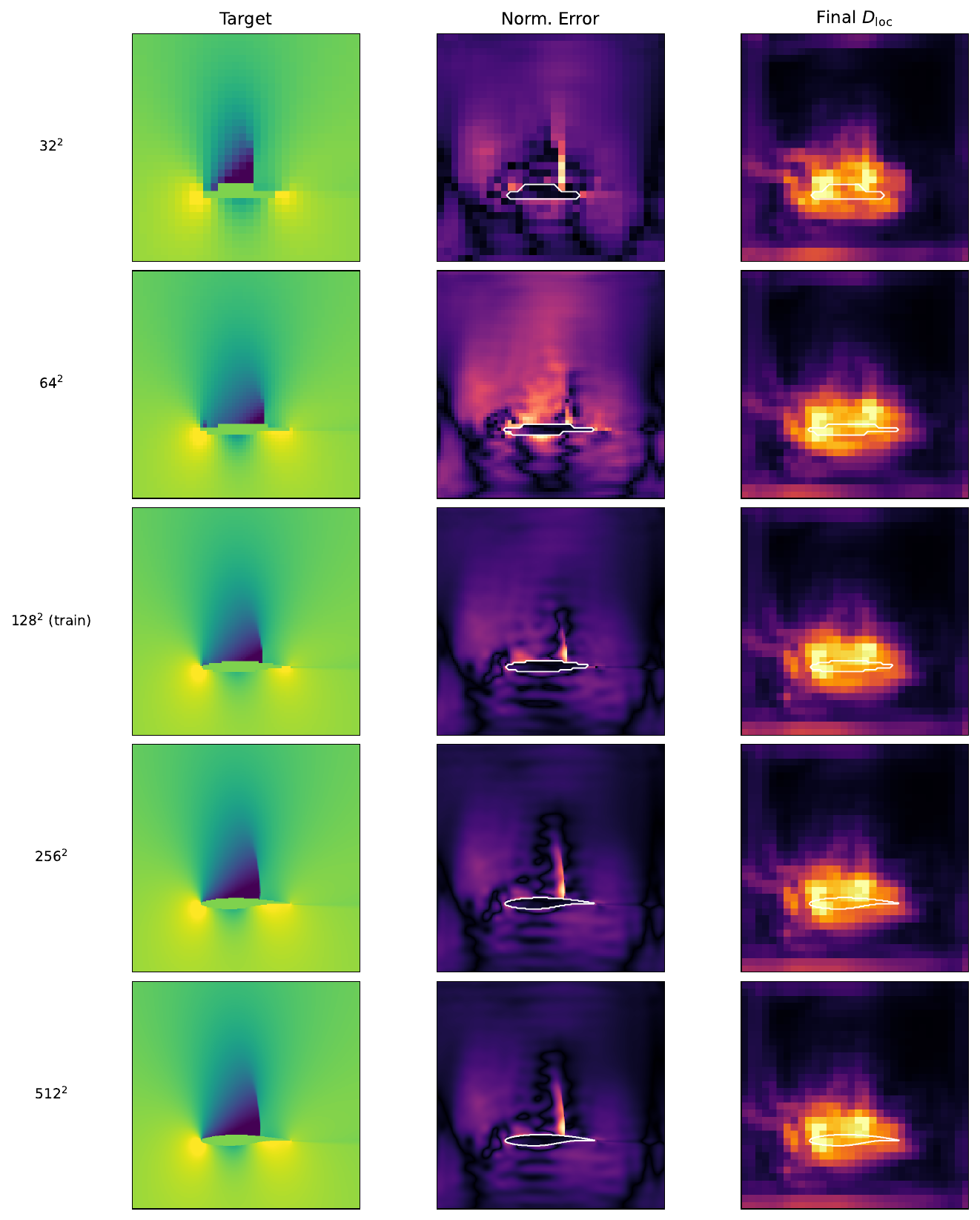}
\end{minipage}
\caption{\textbf{Examples of local incoherence across resolution change.}
Rows correspond to \textit{Continuous Translation}, \textit{Discontinuous
Translation}, and \textit{Airfoil}, and groups within each row show
representative examples at the evaluated resolutions. Within each resolution
group, columns show the target, local absolute error, and final-depth
$D_{\mathrm{loc}}$. Translation panels overlay the transported interface,
while \textit{Airfoil} panels overlay the solid boundary. Error and
$D_{\mathrm{loc}}$ color limits are shared across resolutions within each task
row to make changes in concentration and spatial support visually comparable.
For \textit{Continuous Translation}, $D_{\mathrm{loc}}$ forms a wider and less
sharply peaked halo around the interface off-grid, consistent with the lower
off-grid AP. For \textit{Discontinuous Translation}, the incoherence band
around the jump set becomes thinner and more concentrated at finer
resolutions, consistent with the improving \textbf{AUROC} and \textbf{AP}. For
\textit{Airfoil}, the near-body and wake-associated structure of
$D_{\mathrm{loc}}$ remains visually stable across resolutions, consistent
with the nearly constant \textbf{AUROC} and \textbf{AP}.}
\label{fig:resolution-dloc-samples}
\end{figure}

\subsection{Formation of Localization through Depth}

Figure~\ref{fig:resolution-dloc-depth} shows that the resolution-robust signal
is not inherited from the random initialization. At $r=0$, $D_{\mathrm{loc}}$ has essentially zero correlation with local error
on every task and grid. On the translation tasks, $\rho_E$ then grows through
successive KNO updates along nearly overlapping trajectories across
resolutions. For example, on \textit{Cont. Translation} at $512^2$, it evolves
from approximately zero at initialization to $0.466$, $0.676$, $0.800$, and
$0.923$ at the selected outputs after layers 2, 4, 6, and 8, respectively. The final
localization is therefore constructed by the learned dynamics in a manner
that is largely insensitive to the evaluation grid.

On \textit{Airfoil}, the depthwise evolution is less monotone. Geometry
alignment forms early, while error alignment appears later and varies more
strongly between checkpoints. This matches the distinction observed at final
depth: the latent field consistently identifies structured near-body regions,
but the residual is not governed by a single transported interface.

\begin{figure}[!ht]
\centering
\begin{minipage}[t]{0.325\textwidth}
\centering
\includegraphics[width=\linewidth]{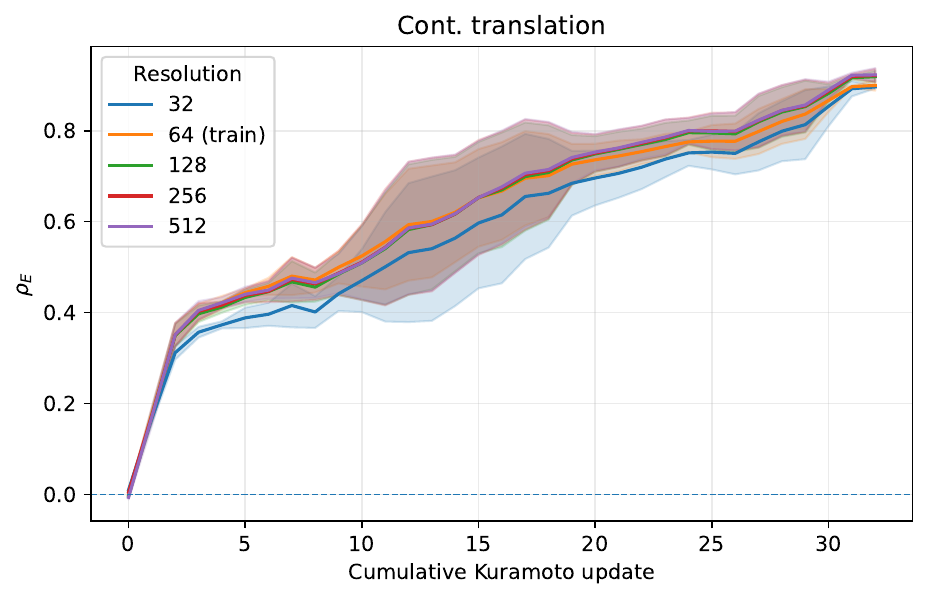}
\end{minipage}
\hfill
\begin{minipage}[t]{0.325\textwidth}
\centering
\includegraphics[width=\linewidth]{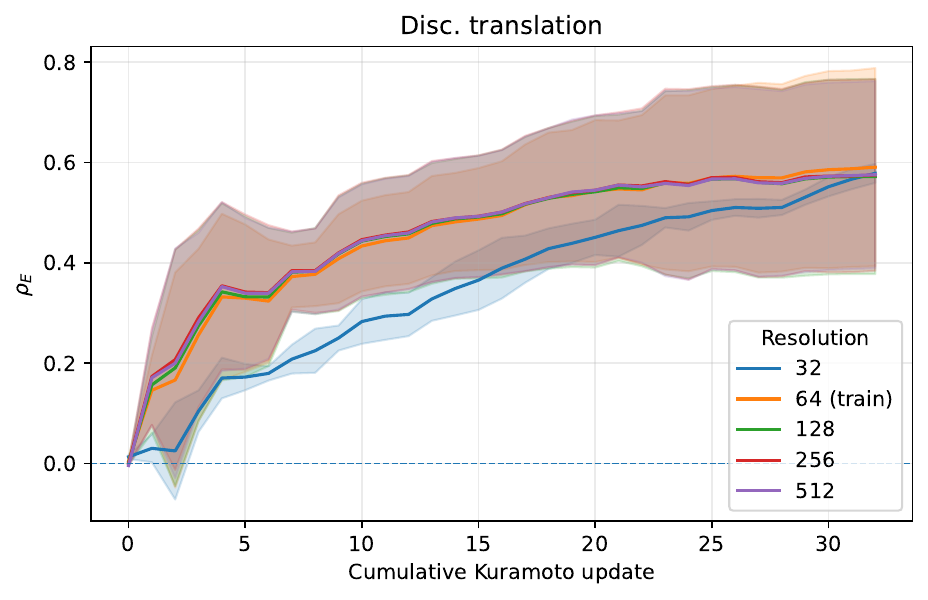}
\end{minipage}
\hfill
\begin{minipage}[t]{0.325\textwidth}
\centering
\includegraphics[width=\linewidth]{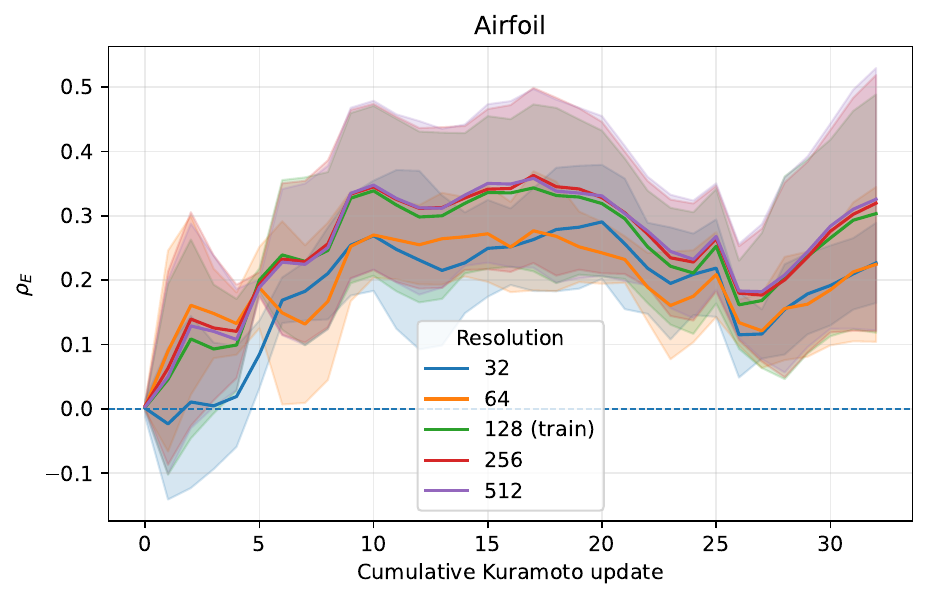}
\end{minipage}
\caption{Depthwise formation of error alignment under resolution transfer.
Curves show the median within-field Spearman correlation between
$D_{\mathrm{loc}}^{(r)}$ and local absolute error; lines and bands are the
mean and standard deviation over checkpoints. The initialization is
uninformative, while task-specific localization emerges through the learned
updates at every resolution.}
\label{fig:resolution-dloc-depth}
\end{figure}

Figure~\ref{fig:resolution-dloc-depth-maps} provides the resolution-transfer
counterpart of the depthwise maps in Figure~\ref{fig:dloc-spatial}.

\begin{figure}[!ht]
\centering
\includegraphics[width=\linewidth]{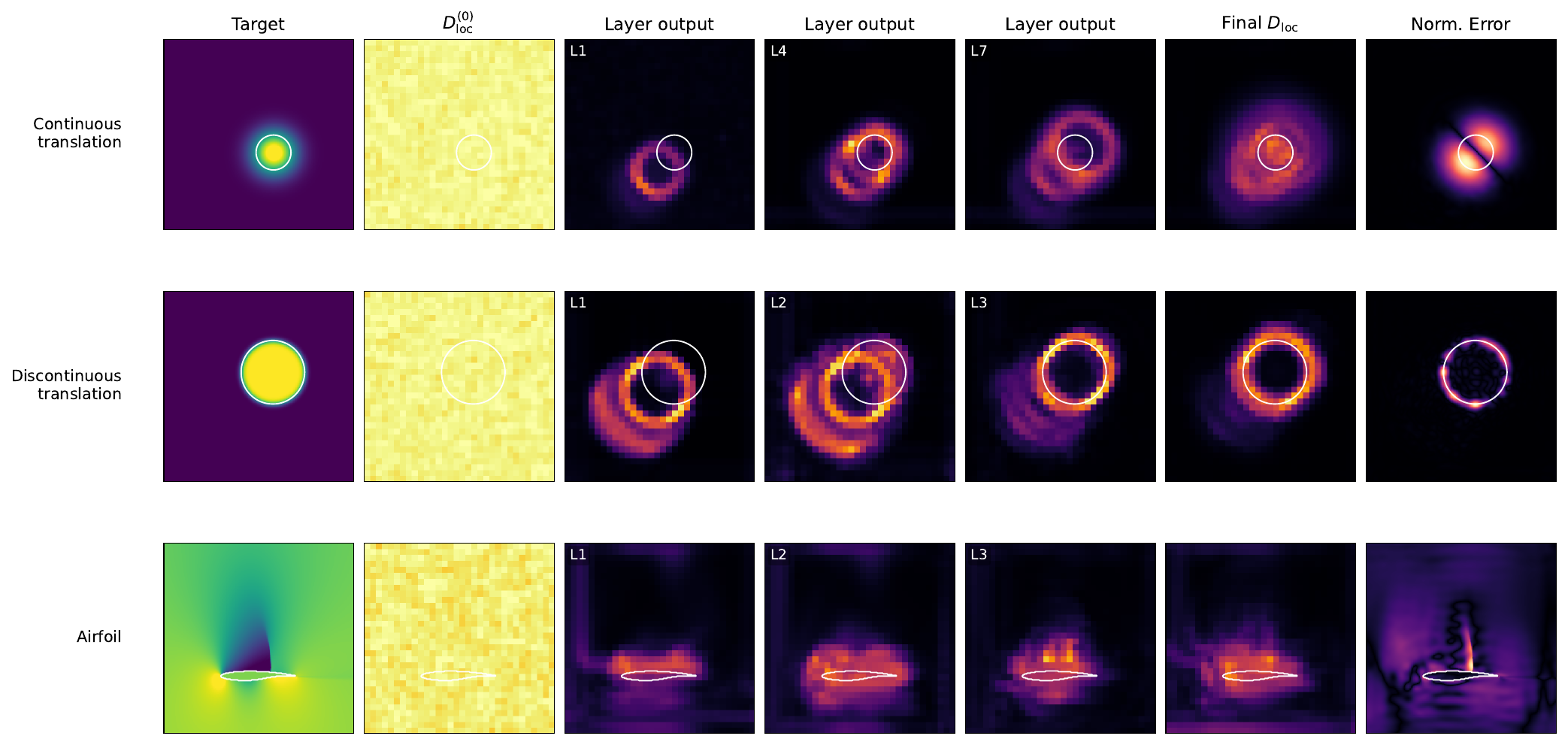}
\caption{\textbf{Formation of local incoherence at} $\mathbf{256^2}.$
Rows show representative samples from \textit{Continuous Translation},
\textit{Discontinuous Translation}, and \textit{Airfoil}. Columns show the
target, local incoherence at initialization $D_{\mathrm{loc}}^{(0)}$,
selected intermediate layer outputs, final $D_{\mathrm{loc}}$, and normalized
local absolute error. Through depth, the initially diffuse incoherence becomes
concentrated around the transported interface on the translation tasks and
around the airfoil surface and nearby flow structures on \textit{Airfoil}.
White contours mark the target interface on the translation tasks and the
solid boundary on \textit{Airfoil}. The $D_{\mathrm{loc}}$ color scale is
shared across depth within each row.}
\label{fig:resolution-dloc-depth-maps}
\end{figure}

\subsection{Interpretation}
This section shows that local incoherence transfers robustly as a
spatial risk ranking across resolutions. Overall, final-depth $D_{\mathrm{loc}}$ acts as a resolution-robust local
error-risk field. Its broad spatial ranking survives zero-shot changes of the
input and output grid without a learned uncertainty head or resolution-specific recalibration. At the same time, the variation of \textbf{AP} and the checkpoint dependence of $\rho_E$ on \textit{Airfoil} show that resolution robustness of localization should not be conflated with universal calibration of error magnitude or of the most extreme residual pixels.

\section{Task Specifications}
\label{app:task-spec}
All experiments use the Representative PDE Benchmark
Dataset~\cite{raonic2023convolutional}. The training, validation, in-distribution (ID) test, and
out-of-distribution (OOD) test splits are fixed before model selection and
shared across all architectures and random seeds.
Table~\ref{tab:benchmarks} summarizes the dataset sizes, resolutions, and
scaling conventions, while Table~\ref{tab:task-generation} specifies the task
generators and OOD shifts.

For \textit{Poisson}, \textit{Wave}, \textit{Allen--Cahn}, \textit{Darcy},
and \textit{Navier--Stokes}, the input and target fields are normalized
independently by fixed dataset-level affine maps to $[0,1]$. For datasets
using affine min--max normalization, the input field $a$ and target field
$u$ are transformed independently according to
\begin{equation*}
\widetilde{a}(x)
=
\frac{a(x)-a_{\min}^{\mathrm{ID}}}
     {a_{\max}^{\mathrm{ID}}-a_{\min}^{\mathrm{ID}}},
\quad
\widetilde{u}(x)
=
\frac{u(x)-u_{\min}^{\mathrm{ID}}}
     {u_{\max}^{\mathrm{ID}}-u_{\min}^{\mathrm{ID}}}.
\end{equation*}
4 scalar normalization constants are read from the corresponding ID
HDF5 archive, and the same constants are reused for the training,
validation, ID-test, and OOD-test splits. \textit{Continuous Translation},
\textit{Discontinuous Translation}, and \textit{Airfoil} are loaded in their stored
scales without additional normalization. For the translation datasets, the
stored fields are already normalized by the data-generation procedure; the
stored \textit{Airfoil} fields remain unnormalized.

Several tasks use random multiscale fields:
\begin{equation*}
f_{K,r}(x,y)
=
\frac{\pi}{K^2}
\sum_{i,j=1}^{K}
a_{ij}(i^2+j^2)^{-r}
\sin(\pi i x)\sin(\pi j y),
\quad
a_{ij}\overset{\mathrm{i.i.d.}}{\sim}\mathcal U(-1,1).
\end{equation*}

\begin{table*}[!ht]
\centering
\caption{Benchmark datasets. Split sizes are reported as
$N_{\mathrm{train}}/N_{\mathrm{validation}}/N_{\mathrm{test\ ID}}/N_{\mathrm{test \ OOD}}$.
``ID affine'' denotes separate fixed affine maps for the input and target,
with constants determined from the ID data and reused for all splits.
``Stored scale'' denotes the values stored in the released HDF5 archives,
without an additional loader-side transformation.}
\label{tab:benchmarks}
\small
\setlength{\tabcolsep}{6pt}
\begin{tabular}{lccc}
\hline
Task & Domain and resolution & Split sizes & Scaling \\
\hline
\textit{Poisson}
& $[0,1]^2$, $64^2$
& $1024/128/256/256$
& ID affine to $[0,1]$ \\
\textit{Wave}
& $[0,1]^2$, $64^2$
& $512/128/256/256$
& ID affine to $[0,1]$ \\
\textit{Allen--Cahn}
& $[0,1]^2$, $64^2$
& $256/128/128/128$
& ID affine to $[0,1]$ \\
\textit{Cont.\ translation}
& $[0,1]^2$, $64^2$
& $512/256/256/256$
& Stored scale \\
\textit{Disc.\ translation}
& $[0,1]^2$, $64^2$
& $512/256/256/256$
& Stored scale \\
\textit{Darcy}
& $[0,1]^2$, $64^2$
& $256/128/128/128$
& ID affine to $[0,1]$ \\
\textit{Navier--Stokes}
& $[0,1]^2$, $64^2 \ (\text{ID}), \ 128^2 \ (\text{OOD})$
& $750/128/128/128$
& ID affine to $[0,1]$ \\
\textit{Airfoil}
& $[-0.75,1.75]^2$, $128^2$
& $512/128/128/128$
& Stored scale \\
\hline
\end{tabular}
\end{table*}

\begin{table*}[!ht]
\centering
\caption{Task definitions and OOD shifts. Parameters not listed in the OOD
column are unchanged from the ID distribution.}
\label{tab:task-generation}
\scriptsize
\setlength{\tabcolsep}{4pt}
\renewcommand{\arraystretch}{1.15}
\begin{tabular}{p{0.12\textwidth}p{0.24\textwidth}p{0.30\textwidth}p{0.25\textwidth}}
\hline
Task & Operator and target & ID distribution & OOD shift \\
\hline
\textit{Poisson}
&
$-\Delta u=f$ in $D$, with $u|_{\partial D}=0$. The target is the closed-form
solution in the sine basis.
&
Source term $f=f_{16,-1/2}$.
&
Increase the spectral cutoff from $K=16$ to $K=20$, introducing forcing
frequencies absent from training.
\\
\textit{Wave}
&
$u_{tt}-c^2\Delta u=0$, with $c=0.1$. The target is the exact solution at
$T=5$.
&
Initial condition $f=f_{24,1}$.
&
Replace $f_{24,1}$ with $f_{32,0.85}$, increasing the number and relative
strength of high-frequency modes.
\\
\textit{Allen--Cahn}
&
$u_t=\Delta u-\varepsilon^2u(u^2-1)$, with $\varepsilon=220$. The target is
$u(\cdot,T)$ at $T=2\times10^{-4}$, computed with an explicit
finite-difference solver.
&
Initial condition $f=f_{24,1}$.
&
Set $K=16$ and sample $r\sim\mathcal U(0.85,1.15)$ independently for each
sample.
\\
\textit{Continuous translation}
&
$u_t+v\cdot\nabla u=0$, with $v=(0.2,0.2)$. The target is
$u(\cdot,1)=f(\cdot-v)$.
&
Radial Gaussian fields with centers sampled from $(0.2,0.4)^2$ and variances
sampled from $(0.003,0.009)$.
&
Shift the center distribution to $(0.4,0.6)^2$.
\\
\textit{Discontinuous translation}
&
$u_t+v\cdot\nabla u=0$, with $v=(0.2,0.2)$. The target is
$u(\cdot,1)=f(\cdot-v)$.
&
Disk indicators with centers sampled from $(0.2,0.4)^2$ and radii sampled
from $(0.1,0.2)$. Samples are generated at $128^2$, smoothed with a Gaussian
filter with $\sigma=1.75$, and band-limited-downsampled to $64^2$.
&
Shift the center distribution to $(0.4,0.6)^2$.
\\
\textit{Darcy}
&
$-\nabla\cdot(a\nabla u)=1$ in $D$, with $u|_{\partial D}=0$. The input is
$a$, and the target is $u$.
&
$a=\psi_{\#}\mu$, where $\mu$ is a mean-zero Gaussian process with
squared-exponential kernel, variance $0.1$, and length scale $\ell=0.1$.
The map $\psi$ sends positive values to $12$ and negative values to $3$.
&
Decrease the Gaussian-process length scale from $\ell=0.1$ to $l=0.05$.
\\
\textit{Navier--Stokes}
&
$\partial_t u+(u\cdot\nabla)u\!=\!-\nabla p+\nu\Delta u$, $\nabla\cdot u\!=\!0$ in $D = \mathbb{T}^2$, with $u = (u_1, u_2)$. The target is $u_1(\cdot, T)$ at $T = 1$, input is $u_1(\cdot, 0)$.
Viscosity $\nu = 4\times10^{-4}$ is applied only to Fourier modes with wavenumber $\ge 12$, modeling flow at very high Reynolds number; no external forcing is applied.
Reference solutions are computed at $128^2$ using a
spectral-viscosity method and then downsampled to $64^2$.
&
The initial condition consists of two oppositely oriented $\tanh$ shear
layers of thickness $\rho=0.1$, centered at $y=0.25$ and $y=0.75$, with
zero initial vertical velocity. Their interfaces are perturbed by
$\sigma_\delta(x)=\delta\sum_{k=1}^{10}
\alpha_k\sin(2\pi kx-\beta_k)$, where
$\delta=0.025$, $\alpha_k\sim\mathcal{U}(0,1)$, and
$\beta_k\sim\mathcal{U}(0,2\pi)$.
&
Set $\rho=0.09$ and move the shear-layer centers to $y=0.3$ and $y=0.7$.
Evaluation is performed directly at $128^2$. Thus the split combines a
parameter shift with zero-shot resolution transfer.
\\
\textit{Airfoil}
&
The operator maps the characteristic function of a perturbed RAE2822 airfoil
to the steady-state density of compressible Euler flow with $\gamma=1.4$.
Solutions are computed on a $243\times43$ elliptic mesh and interpolated to
$128^2$; errors are evaluated only outside the airfoil.
&
Shapes use $20$ Hicks--Henne bump functions with coefficients sampled from
$[0,1]$. The freestream conditions are $M_\infty=0.729$,
$\alpha=2.31^\circ$, and $T_\infty=p_\infty=1$.
&
Increase the number of Hicks--Henne bump functions from $20$ to $30$.
\\
\hline
\end{tabular}
\end{table*}

\section{Baseline Architecture Details}
\label{app:baseline-architectures}

This section describes the implementation of each baseline architecture. All
models map one scalar input field to one scalar output field. The training
resolution is $64^2$ for all tasks except \textit{Airfoil}, which uses
$128^2$. Details of the hyperparameter optimization procedure and the selected configurations are given in Appendix~\ref{app:training-protocol}.

\subsection{Fourier Neural Operator}

We use the FNO implementation from the \texttt{neuraloperator} library
\cite{li2020fourier}. The input is augmented internally with two Cartesian
coordinate channels and lifted pointwise to width $w$. The network contains
$L$ Fourier layers retaining $m$ modes along each spatial axis. The lifting
and projection channel ratios are fixed to two. Domain padding of width $p$
is applied on the right and bottom and removed after the output projection.

\subsection{Weight-Tied FNO and FNO-DEQ}

FNO-WT and FNO-DEQ \cite{marwah2023deep} use input lifting, shared Fourier update, and
output projection. The input and coordinate channels are lifted using two
pointwise convolutions $(c_{\mathrm{in}}+2)\rightarrow2w\rightarrow w$,
with a GELU activation between them. The shared update contains $D$
sublayers. Each sublayer applies a spectral convolution, a pointwise linear
map of the current state, and a pointwise injection of the lifted input:
\begin{equation*}
h_{j+1}
=
\operatorname{GELU}
\left(
\mathcal K_jh_j+W_jh_j+B_js
\right),
\quad
j=0,\ldots,D-1,
\end{equation*}
where $s$ denotes the lifted input. The output projection has channel
dimensions $w\rightarrow2w\rightarrow c_{\mathrm{out}}$.

FNO-WT applies the shared update $4$ times and differentiates through the
complete unrolled computation. FNO-DEQ instead solves $z^\star=\Phi_\theta(z^\star;s)$.

The first $300$ optimization steps use eight explicit unrolls. Subsequent
steps use a no-gradient forward equilibrium solve followed by
phantom-gradient backpropagation \cite{geng2021training}. Anderson acceleration \cite{anderson1965iterative} uses history size
$5$, regularization $10^{-4}$, tolerance $10^{-3}$, and mixing coefficient
$1$. The maximum number of forward iterations is $32$, except on
\textit{Navier--Stokes}, where it is $16$. The phantom gradient uses one step
with damping $0.5$, except on \textit{Navier--Stokes}, where it uses three
steps with damping $0.8$.

\subsection{Riesz Neural Operator}

RNO follows the supplementary implementation of the Riesz Neural Operator
\cite{yangriesz}. Cartesian coordinates
are concatenated to the input before pointwise lifting. Each spectral layer
forms
\begin{equation*}
\widehat h_{\mathrm R}
=
\alpha_0\widehat h
+
\alpha_1
\frac{\mathrm i\xi_1}{\lVert\xi\rVert_2}\widehat h
+
\alpha_2
\frac{\mathrm i\xi_2}{\lVert\xi\rVert_2}\widehat h,
\end{equation*}
where the three coefficients are learned and initialized to $1/3$. Learned
complex channel maps are applied to the retained positive and negative
frequency bands, and each spectral branch is combined with a pointwise
linear branch.

For base width $w$, the channel widths are $(\nicefrac{w}{2}, \nicefrac{3w}{4}, w, w, \nicefrac{5w}{4})$,
and $4$ Riesz layers retain $(m, \lfloor\nicefrac{3m}{4}\rfloor, \lfloor\nicefrac{m}{2}\rfloor,\lfloor\nicefrac{m}{2}\rfloor)$
modes. SELU is applied after the first three layers. Two additional
SELU-activated pointwise layers and a linear output map process the final
feature field.

\subsection{ConvFNO}

ConvFNO follows the local--spectral construction of
\cite{liu2025enhancing}. The scalar input and two coordinate channels are
processed by a three-level UNet local branch. For base width
$c_{\mathrm U}$, its encoder widths are $c_{\mathrm U}$, $2c_{\mathrm U}$, $4c_{\mathrm U}$, $8c_{\mathrm U}$.
Each local block contains two circularly padded $3\times3$ convolutions with
ReLU activations. The decoder uses transposed convolutions and skip
connections. A final $1\times1$ convolution produces
$c_{\mathrm{LSF}}$ local features.

The local features are concatenated with the original input and coordinate
channels and passed to a $4$-layer FNO of width $w$. Its lifting and
projection widths are $r_{\mathrm{lift}}w$ and $r_{\mathrm{proj}}w$,
respectively. The spectral branch uses group normalization, channel MLPs,
linear residual paths, and Tucker-factorized spectral weights with rank
parameter $1$. Fourier-domain padding is zero. The local branch is evaluated
at the training resolution; during resolution-transfer evaluation, its output
is interpolated to the requested grid before concatenation with the
native-resolution input.

\subsection{Convolutional Neural Operator}

CNO follows the encoder--decoder architecture of
\cite{raonic2023convolutional}. A CNO block applies a $3\times3$
convolution, batch normalization, and the alias-controlled CNO activation.
The activation bicubically upsamples the field by a factor of two, applies
LeakyReLU, and bicubically resamples the result to the requested output
resolution. The lifting and projection blocks use intermediate width $64$.
The decoder uses encoder skip connections and the additional invariant blocks
from the reference implementation.

An architecture is specified by $(L,R,R_{\mathrm{neck}},d_e)$,
where $L$ is the number of resolution levels, $R$ is the number of residual
blocks at each encoder level, $R_{\mathrm{neck}}$ is the number of residual
blocks at the bottleneck, and $d_e$ is the channel multiplier. The encoder
widths are $\nicefrac{d_e}{2}$, $d_e$, $\ldots$, $2^{L-1}d_e$.
All configurations use batch normalization, $3\times3$ kernels, the
additional invariant decoder blocks, and lifting/projection intermediate
width $64$. Inputs evaluated away from the training resolution are
spectrally resized to the training grid, and predictions are resized back to
the grid with requested resolution.

\subsection{DeepONet}

We use the Cartesian-product DeepONet implementation from DeepXDE
\cite{lu2021learning}. The branch network receives the complete input field
flattened at the training resolution. Inputs on another grid are bilinearly
interpolated to the training resolution before entering the branch network.
The trunk network receives the coordinate $(x,y)$ of each requested output
point. For basis size $b$, the output is
\begin{equation*}
\widehat u(x)
=
\sum_{j=1}^{b}B_j(a)T_j(x)+\beta.
\end{equation*}
All linear layers use Glorot-normal initialization. An architecture is
specified by $(b,w_b,d_b,w_t,d_t,\sigma)$, where $w_b,d_b$ and $w_t,d_t$
denote the branch and trunk widths and depths, and $\sigma$ is the shared
activation.

\subsection{UNet}

The UNet baseline uses $4$ encoder levels with widths $c$, $2c$, $4c$, $8c$, and a bottleneck of width $16c$. Each block contains two $3\times3$
convolutions, each followed by batch normalization and $\tanh$. Downsampling
uses $2\times2$ max pooling. Upsampling uses $2\times2$ transposed
convolutions followed by concatenation with the corresponding encoder feature
map. A final $1\times1$ convolution produces the output. Inputs whose
dimensions are not divisible by $16$ are replication-padded on the right and
bottom and cropped after the output projection.

\subsection{AKOrN}

AKOrN denotes our resolution-preserving adaptation of the original
architecture \cite{miyato2025artificial} to dense operator learning. We
remove all spatial downsampling and upsampling operations and replace the
original task-specific prediction head with a pointwise field projection.
The oscillator and conditioning fields therefore remain on the input grid
throughout the network.

A zero-padded $3\times3$ convolution lifts the scalar input to a conditioning
field of width $w=64$. The oscillator field has the same width and is grouped
into $M=16$ oscillators of dimension $n=4$ at every spatial location. Initial
oscillator vectors are sampled from a standard Gaussian and normalized
oscillator-wise.

We use the same layer and integration-step indices as in
Section~\ref{sec:methodology}. AKOrN has layers
$l=0,\ldots,L-1$ with $L=3$, and each layer performs
$r=0,\ldots,R-1$ Euler--normalization steps with $R=4$. Let $q_l^{(r)}$
and $c_l$ denote the oscillator and conditioning fields in layer $l$. The
update is
\begin{align}
g_l^{(r)}
&=
\operatorname{Proj}_{q_l^{(r)}}
\left(
\mathcal C_{k_l}\!\left(q_l^{(r)}\right)
+
c_l
\right)
+
\Omega q_l^{(r)},
\nonumber\\
q_l^{(r+1)}
&=
\frac{
q_l^{(r)}+\gamma g_l^{(r)}
}{
\left\lVert
q_l^{(r)}+\gamma g_l^{(r)}
\right\rVert_2
},
\quad
r=0,\ldots,R-1.
\end{align}
The step size is fixed at $\gamma=1$. The natural-frequency term is a learned
global planar rotation initialized with magnitude one and shared across
spatial positions and oscillator channels.

After each layer, a $3\times3$ invariant convolution maps the oscillator
field to $64\times2$ channels. These channels are grouped in pairs and
reduced by their Euclidean norm, producing a $64$-channel invariant field.
Two residual pre-activation convolutions and one additional pre-activation
convolution process this field to form $c_{l+1}$. Batch normalization is used
in the readout head. A final $1\times1$ convolution maps $c_L$ to the scalar
output. 

Since AKOrN is not inherently a neural operator baseline, its hyperparameters are configured once as 
\begin{equation*}
n=4,
\quad
M=16,
\quad
w=64,
\quad
L=3,
\quad
R=4,
\end{equation*}
\begin{equation*}
(k_0,k_1,k_2)=(9,7,5),
\quad
k_{\mathrm{read}}=3,
\quad
r_{\mathrm{read}}=2.
\end{equation*}
and kept fixed across all tasks.

\section{KNO Architecture and Implementation Details}
\label{app:implementation-details}
\label{app:additional-methodology}

\subsection{Forward-Pass Algorithm}
\label{app:kno-algorithm}

Let $a\in\mathbb R^{B\times c_{\mathrm{in}}\times H\times W},
\ S=(H,W),\ S_c=(H_c,W_c)$ denote an input batch, its grid, and the canonical grid. All benchmark configurations reported in Table~\ref{tab:main-results} use $c_{\mathrm{in}}=c_{\mathrm{out}}=1,  S_c=(32,32), \ c=64, \ M=16, \ n=4, \ R=4$.

We use exactly the indexing of Section~\ref{sec:methodology}: layers are
indexed by $l=0,\ldots,L-1$, stages in layer $l$ by
$t=0,\ldots,T_l-1$, and weight-tied integration steps by
$r=0,\ldots,R-1$. KNO maintains $h_l\in
\mathbb R^{B\times c\times H_l\times W_l}, \
q_{l,t}^{(r)}
\in
\mathbb R^{B\times M\times n\times H_c\times W_c}, \
s_{l,t}
\in
\mathbb R^{B\times Mn\times H_c\times W_c}$.
At a stage boundary, $q_{l,t}=q_{l,t}^{(0)}$, and after the stage
$q_{l,t+1}=q_{l,t}^{(R)}$.

We write $\mathcal R_{S\rightarrow S'}$ for centered Fourier resampling from
grid $S$ to grid $S'$. It crops centered Fourier coefficients when reducing
resolution and zero-pads the centered spectrum when increasing it. The
low-pass operator $\Pi_{\leq k}$ retains a centered frequency window;
all reported configurations use $k=16$.

The input is augmented with normalized Cartesian coordinates $\xi$, lifted
to feature width $c$, and band-limited onto the canonical grid:
\begin{equation*}
h_0=\mathrm{MLP_1}([a;\xi]),
\quad
\bar h_0
=
\mathcal R_{S\rightarrow S_c}
\left(
\Pi_{\leq k}h_0
\right).
\end{equation*}
At every forward pass, the initial oscillator field is sampled from a
standard Gaussian and normalized oscillator-wise:
\begin{equation*}
\varepsilon_m(x)\sim\mathcal N(0,I_n),
\quad
q_{0,0,m}^{(0)}(x)
=
\frac{\varepsilon_m(x)}
{\lVert\varepsilon_m(x)\rVert_2},
\quad
m=1,\ldots,M.
\end{equation*}
The initial stimulus is $s_{0,0}=\mathrm{MLP_2}(\bar h_0)$.

At layer $l$, the feature field is transferred once to the canonical grid,
\begin{equation*}
h_l^c
=
\mathcal R_{S_l\rightarrow S_c}[h_l],
\end{equation*}
and held fixed throughout the stages of that layer. At stage $t$ and
integration step $r$, the local-message network produces
\begin{equation*}
c_{l,t}^{(r)}
=
\mathcal C_{l,t}
\left(
q_{l,t}^{(r)},h_l^c
\right).
\end{equation*}
The unconstrained force on oscillator $m$ is
\begin{equation*}
F_{l,t,m}^{(r)}
=
c_{l,t,m}^{(r)}
+
s_{l,t,m}
+
\Omega_{l,t,m}q_{l,t,m}^{(r)},
\end{equation*}
where $\Omega_{l,t,m}$ is skew-symmetric. With
\begin{equation*}
\operatorname{Proj}_{q}(F)
=
F-\langle F,q\rangle q,
\end{equation*}
one integration step is
\begin{equation*}
q_{l,t,m}^{(r+1)}
=
\frac{
q_{l,t,m}^{(r)}
+
\gamma_{l,t}
\operatorname{Proj}_{q_{l,t,m}^{(r)}}
\left(
F_{l,t,m}^{(r)}
\right)
}{
\left\lVert
q_{l,t,m}^{(r)}
+
\gamma_{l,t}
\operatorname{Proj}_{q_{l,t,m}^{(r)}}
\left(
F_{l,t,m}^{(r)}
\right)
\right\rVert_2
}.
\end{equation*}
The local-message parameters, skew-symmetric map, and learned scalar
$\gamma_{l,t}$ are shared across the $R$ integration steps of the stage,
while the stimulus $s_{l,t}$ is held fixed. After the final step,
\begin{equation*}
q_{l,t+1}=q_{l,t}^{(R)}.
\end{equation*}

The invariant readout then refreshes the stimulus:
\begin{equation*}
\widetilde s_{l,t+1}
=
g_{l,t}\!\left(q_{l,t+1}\right),
\quad
\alpha_l=\sigma(a_l),
\end{equation*}
\begin{equation*}
s_{l,t+1}
=
\alpha_l s_{l,t}
+
(1-\alpha_l)\widetilde s_{l,t+1}.
\end{equation*}
The retention coefficient $\alpha_l$ is shared across the stages of layer
$l$.

After all $T_l$ stages, the terminal oscillator state is decoded and added
residually to the feature field:
\begin{equation*}
h_{l+1}
=
\mathcal R_{S_l\rightarrow S_{l+1}}[h_l]
+
\mathcal R_{S_c\rightarrow S_{l+1}}
\left[
D_l\!\left(q_{l,T_l}\right)
\right].
\end{equation*}
For fixed-resolution evaluation, $S_{l+1}=S_l$. When another output
resolution is requested, only the final layer targets that grid. The
oscillator and stimulus states are carried between layers as
\begin{equation*}
q_{l+1,0}=q_{l,T_l},
\quad
s_{l+1,0}=s_{l,T_l}.
\end{equation*}
The prediction is $\widehat u=\mathrm{MLP_3}(h_L)$.

Algorithm~\ref{alg:kno-forward} summarizes the complete forward pass.

\begin{algorithm}{Kuramoto Neural Operator forward pass}
\label{alg:kno-forward}
\begin{algorithmic}[1]
\Require Input field $a$, stage schedule $(T_0,\ldots,T_{L-1})$,
integration-step count $R$
\Ensure Predicted field $\widehat u$
\State Append normalized coordinates and compute
$h_0\leftarrow P([a;\xi])$
\State Compute
$\bar h_0\leftarrow
\mathcal R_{S\rightarrow S_c}(\Pi_{\leq k} h_0)$
\State Sample and normalize Gaussian oscillator vectors to obtain $q_{0,0}$
\State Compute $s_{0,0}\leftarrow E(\bar h_0)$
\For{$l=0,\ldots,L-1$}
    \State Compute
    $h_l^c\leftarrow\mathcal R_{S_l\rightarrow S_c}(h_l)$
    \State Compute $\alpha_l\leftarrow\sigma(a_l)$
    \For{$t=0,\ldots,T_l-1$}
        \State Set $q_{l,t}^{(0)}\leftarrow q_{l,t}$
        \For{$r=0,\ldots,R-1$}
            \State
            $F\leftarrow
            \mathcal C_{l,t}(q_{l,t}^{(r)},h_l^c)
            +s_{l,t}+\Omega_{l,t}q_{l,t}^{(r)}$
            \State
            $q_{l,t}^{(r+1)}
            \leftarrow
            \operatorname{Normalize}
            \left(
            q_{l,t}^{(r)}
            +
            \gamma_{l,t}
            \operatorname{Proj}_{q_{l,t}^{(r)}}(F)
            \right)$
        \EndFor
        \State $q_{l,t+1}\leftarrow q_{l,t}^{(R)}$
        \State
        $s_{l,t+1}\leftarrow
        \alpha_l s_{l,t}
        +(1-\alpha_l)g_{l,t}(q_{l,t+1})$
    \EndFor
    \State
    $h_{l+1}\leftarrow
    \mathcal R_{S_l\rightarrow S_{l+1}}(h_l)
    +
    \mathcal R_{S_c\rightarrow S_{l+1}}
    \left(
    D_l(q_{l,T_l})
    \right)$
    \State Set
    $q_{l+1,0}\leftarrow q_{l,T_l}$ and
    $s_{l+1,0}\leftarrow s_{l,T_l}$
\EndFor
\State $\widehat u\leftarrow Q(h_L)$
\Return $\widehat u$
\end{algorithmic}
\end{algorithm}

\subsection{Layer Specifications}
\label{app:kno-layer-specifications}
Table~\ref{tab:kno-layer-specifications} summarizes the component-level
channel dimensions and spatial operations used by KNO.
All spatial convolutions in the local-message, stimulus-encoder, and
stimulus-readout networks use circular padding. Pointwise maps use
$1\times1$ convolutions. The lifting, layer decoder, and output projection are
two-layer \texttt{ChannelMLP} modules. Let $D_q=Mn$ denote the number of flattened oscillator channels.

\paragraph{Grid embedding and lifting.}
Two normalized Cartesian coordinate channels are appended to the scalar
input before a two-layer pointwise lifting.

\paragraph{Initial spectral projection.}
The lifted feature field is low-pass filtered and spectrally transferred to
the canonical $32\times32$ grid. This operation contains no learned
parameters.

\paragraph{Stimulus encoder.}
Let $H_s=\max\!\left(c,D_q,\operatorname{round}(\rho_s c)\right)$, where $\rho_s=1.5$ in all reported configurations. Three circular $3\times3$ convolutions with channel dimensions $c\rightarrow H_s\rightarrow H_s\rightarrow D_q$ and GELU activations map the canonical feature field to the initial stimulus.

\paragraph{Local-message network.}
The oscillator and canonical feature fields are concatenated. A local branch
applies $5\times5$ and $3\times3$ convolutions, each followed by GELU. The
result is concatenated with the original input and passed through a pointwise
channel MLP that produces the $D_q$-channel interaction force. Every stage
has an independent local-message network.

\paragraph{Oscillator dynamics.}
Each stage has one learned scalar $\gamma_{l,t}$, one skew-symmetric rotation
map, and one local-message network, all shared across its $R$ integration
steps.

\paragraph{Stimulus readout.}
The oscillator field is expanded pointwise from $D_q$ to $D_q n$ channels,
reshaped into $D_q$ groups of dimension $n$, and reduced by the Euclidean
norm. A learned bias and a lightweight convolutional head then produce the
next stimulus candidate. Every stage has an independent readout.

\paragraph{Layer decoder and output projection.}
Each layer has a separate pointwise decoder from the terminal oscillator
state to a $64$-channel feature correction. The correction is spectrally
transferred to the feature grid and added residually. A final pointwise
projection produces the scalar output field.

\begin{table*}[!ht]
\centering
\caption{KNO layer specifications shared by the reported configurations.}
\label{tab:kno-layer-specifications}
\small
\begin{tabular}{lll}
\hline
Component & Channel dimensions & Spatial operations \\
\hline
Grid embedding
& $1\rightarrow3$
& Append two normalized coordinates \\
Lifting $P$
& $3\rightarrow128\rightarrow64$
& Pointwise \texttt{ChannelMLP} \\
Initial projection
& $64\rightarrow64$
& Low-pass filter and Fourier resampling to $32^2$ \\
Stimulus encoder $E$
& $64\rightarrow H_s\rightarrow H_s\rightarrow D_q$
& Three circular $3\times3$ convolutions \\
Local convolution
& $(64+D_q)\rightarrow32\rightarrow32$
& Circular $5\times5$ and $3\times3$ convolutions \\
Local readout
& $(96+D_q)\rightarrow(96+D_q)\rightarrow D_q$
& Pointwise \texttt{ChannelMLP} \\
Invariant expansion
& $D_q\rightarrow D_q n$
& Pointwise $1\times1$ convolution \\
Invariant reduction
& $D_qn\rightarrow D_q$
& Norm over $D_q$ groups of dimension $n$ \\
Stimulus post-processing
& $D_q\rightarrow D_q\rightarrow D_q$
& Circular $3\times3$ and pointwise $1\times1$ convolutions \\
Layer decoder $D_l$
& $D_q\rightarrow\max(64,\lceil1.5D_q\rceil)\rightarrow64$
& Pointwise \texttt{ChannelMLP} and Fourier resampling \\
Output projection $Q$
& $64\rightarrow128\rightarrow1$
& Pointwise \texttt{ChannelMLP} \\
\hline
\end{tabular}
\end{table*}

\subsection{Direct Comparison between KNO and AKOrN}
\label{app:kno-akorn-comparison}

KNO and AKOrN share the same basic dynamical primitive, but embed it into
different surrounding architectures, as summarized in
Table~\ref{tab:kno-akorn-comparison}.

\begin{table*}[!ht]
\centering
\caption{Implementation-level comparison between KNO and the
operator-learning adaptation of AKOrN.}
\label{tab:kno-akorn-comparison}
\small
\resizebox{\linewidth}{!}{%
\begin{tabular}{lll}
\hline
Component & KNO & AKOrN adaptation \\
\hline
Persistent states
& Feature, oscillator, and stimulus fields
& Oscillator and conditioning fields \\
Oscillator grid
& Fixed canonical grid
& Native input grid \\
Coordinate channels
& Appended before pointwise lifting
& Not appended \\
Conditioning of coupling
& Concatenated with $q$ before the message network
& No separate coupling conditioning \\
Local interaction
& $5\times5$ and $3\times3$ convolutions plus channel MLP
& One layer-specific convolution \\
Step size
& Learned separately for each stage
& Fixed globally at $\gamma=1$ \\
Rotation
& Stage- and oscillator-specific skew-symmetric map
& Global planar rotation shared across channels \\
Persistent feedback
& Stimulus EMA across stages and layers
& No separate persistent stimulus \\
Feature update
& Residual decoder from $q$ to a separate feature pathway
& No separate residual feature pathway \\
Resolution handling
& Spectral transfer to and from the canonical grid
& Dynamics evaluated directly on the requested grid \\
Output
& Pointwise projection of the final feature field
& $1\times1$ projection of the final conditioning field \\
\hline
\end{tabular}%
}
\end{table*}

The central difference is the separation of state roles in KNO. The feature
field $h$ stores the evolving operator representation, the oscillator field
$q$ carries the constrained dynamics, and the stimulus $s$ provides
persistent feedback. AKOrN instead alternates between an oscillator field and
a single conditioning field: the condition is added to the oscillator update
and then replaced by the invariant readout after every layer.

KNO also conditions the interaction law itself. Its local-message network
receives the concatenation of $q$ and the canonical-grid feature field,
allowing the current operator representation to modulate the learned
coupling. In AKOrN, the connectivity convolution depends only on $q$, while
the conditioning field enters only as an additive drive. Finally, KNO decodes each
terminal oscillator state as a residual correction to a separate feature
pathway, whereas the adapted AKOrN passes the invariant oscillator readout
directly to the next layer.

\subsection{Architecture Search Space}
\label{app:kno-search-space}

The KNO candidate pool contains KNO-1, KNO-2, and KNO-3. All three
configurations use $k=16,\ c=64,\ M=16,\ n=4,\ R=4,\ S_c=(32,32)$.
Each candidate contains eight stages and therefore $8R=32$ oscillator
integration steps in total. They differ only in the number of feature-update
layers and in the allocation of stages across those layers.

\begin{table*}[!ht]
\centering
\caption{Predeclared KNO architecture candidate pool. The stage schedule is
$(T_0,\ldots,T_{L-1})$.}
\label{tab:kno-search-space}
\small
\begin{tabular}{lccccccc}
\hline
Configuration
& Modes
& Width
& Layers
& Oscillators
& Dimension
& Stage schedule
& Steps $R$ \\
\hline
KNO-1
& $16$ & $64$ & $4$ & $16$ & $4$
& $(2,2,2,2)$
& $4$ \\
KNO-2
& $16$ & $64$ & $6$ & $16$ & $4$
& $(2,1,1,1,1,2)$
& $4$ \\
KNO-3
& $16$ & $64$ & $8$ & $16$ & $4$
& $(1,1,1,1,1,1,1,1)$
& $4$ \\
\hline
\end{tabular}
\end{table*}

The final configurations selected for each task is reported in
Table~\ref{tab:baseline-selected-architectures}.

\section{Training, Validation, and Hyperparameter-Selection Details}
\label{app:training-protocol}

\subsection{Architecture Selection Protocol}
\label{sec:hp-selection}

Every searched model family is selected independently for each task from a
fixed, predeclared candidate pool using the same two-stage, validation-only
procedure. Baseline families use five candidates, while KNO uses the three
candidates defined in Appendix~\ref{app:kno-search-space}.  AKOrN was not originally an operator learning architecture and is used only as a direct adaptation of a closely related architecture, taken as a reference; consequently, a single fixed set of hyperparameters is used across all tasks. Neither ID-test nor OOD-test errors enter architecture, hyperparameter, checkpoint,
or seed selection.

\paragraph{Two-stage selection procedure}

Let $b$ index a searched model family, $d$ a task, and
$\mathcal A_{b,d}$ its candidate set. Each candidate
$\lambda\in\mathcal A_{b,d}$ is first trained for $500$ epochs with seed $0$. Its
stage-one score is
\begin{equation*}
V_{b,d}(\lambda,0)
=
\min_{1\leq e\leq500}
\mathcal E
\left(
\theta_{b,d,\lambda,0}^{(e)};
\mathcal D_d^{\mathrm{val}}
\right),
\end{equation*}
where $\mathcal{E}$ is defined in \eqref{eq:relative-l2}.
A run is eligible only if it reaches epoch $500$. The two candidates with the
lowest stage-one scores form the finalist set $\mathcal F_{b,d}$.

Each finalist is then trained from scratch with seeds $1$ and $2$. Together
with the reused seed-$0$ run, each finalist is represented by three seeds.
Its stage-two score is
\begin{equation*}
V_{b,d}(\lambda)
=
\frac{1}{3}
\sum_{j=0}^{2}
\min_{1\leq e\leq500}
\mathcal E
\left(
\theta_{b,d,\lambda,j}^{(e)};
\mathcal D_d^{\mathrm{val}}
\right),
\end{equation*}
and the selected architecture is
\begin{equation*}
\lambda_{b,d}^{\star}
=
\operatorname*{arg\,min}_{\lambda\in\mathcal F_{b,d}}
V_{b,d}(\lambda).
\end{equation*}

For a family with $N_{\text{cand}}$ candidates, this procedure requires $N_{\text{cand}}+4$ training
runs per task: $N_{\text{cand}}$ seed-$0$ screening runs and $4$ additional runs for the
two finalists. Each five-candidate baseline family therefore uses nine runs,
while the three-candidate KNO pool uses seven. AKOrN uses one fixed
architecture trained with three seeds. The validation criterion is identical
across searched families, although the candidate-pool sizes differ.

For each seed of the selected architecture, the checkpoint with the lowest
validation relative $L_2$ is retained for final evaluation. The selected
seed-$0$, seed-$1$, and seed-$2$ runs are used directly; no additional
post-selection retraining is performed. During training and selection,
ID-test and OOD-test data are not loaded.

\subsection{Baseline Architecture Search Spaces}
\label{app:baseline-search-spaces}

Every searched baseline family uses five predeclared candidates per task.
Table~\ref{tab:baseline-search-spaces} collects the complete candidate pools.
The vectors are ordered as follows: FNO $(m,w,L,p)$; FNO-WT and FNO-DEQ
$(m,w,D,p)$; RNO $(m,w)$; ConvFNO
$(m,w,c_{\mathrm U},c_{\mathrm{LSF}},r_{\mathrm{lift}},r_{\mathrm{proj}})$;
CNO $(L,R,R_{\mathrm{neck}},d_e)$; DeepONet
$(b,w_b,d_b,w_t,d_t,\sigma)$; and UNet $(c)$. FNO-WT and FNO-DEQ share the
same architecture pool and differ only in their iteration and differentiation
procedures. AKOrN uses the single fixed architecture
described in Appendix~\ref{app:baseline-architectures}.

\begin{table*}[!ht]
\centering
\caption{Predeclared baseline architecture candidate pools. Each row lists
all five candidates used for the indicated family and task group.}
\label{tab:baseline-search-spaces}
\scriptsize
\setlength{\tabcolsep}{3.5pt}
\renewcommand{\arraystretch}{1.12}

\resizebox{\linewidth}{!}{%
\begin{tabular}{lll}
\hline
Family & Tasks & Candidate vectors \\
\hline

FNO
& Non-\textit{Airfoil}
& $(12,48,4,8)$, $(16,64,4,8)$, $(16,128,5,0)$, $(20,64,4,8)$, $(20,96,5,0)$ \\
& \textit{Airfoil}
& $(16,48,4,12)$, $(16,64,4,12)$, $(16,128,5,0)$, $(24,64,4,12)$, $(24,96,4,12)$ \\
\hline

FNO-WT/FNO-DEQ
& Non-\textit{Airfoil}
& $(8,40,3,8)$, $(12,32,3,8)$, $(16,24,3,8)$, $(16,32,2,8)$, $(12,32,3,0)$ \\
& \textit{Airfoil}
& $(8,40,3,12)$, $(12,32,3,12)$, $(16,24,3,12)$, $(16,32,2,12)$, $(12,32,3,0)$ \\
\hline

RNO
& Non-\textit{Airfoil}
& $(12,64)$, $(16,96)$, $(20,128)$, $(24,128)$, $(20,160)$ \\
& \textit{Airfoil}
& $(12,64)$, $(16,96)$, $(20,128)$, $(28,128)$, $(20,160)$ \\
\hline

ConvFNO
& \textit{Poisson}, \textit{Wave}, translations, \textit{Darcy}
& $(32,64,16,32,2,2)$, $(24,64,16,32,2,2)$, $(24,48,16,32,2,2)$, $(32,80,16,32,2,2)$, $(32,64,24,48,2,2)$ \\
& \textit{Allen--Cahn}
& $(48,64,16,32,4,2)$, $(32,64,16,32,4,2)$, $(32,48,16,32,4,2)$, $(48,80,16,32,4,2)$, $(48,64,24,48,4,2)$ \\
& \textit{Navier--Stokes}, \textit{Airfoil}
& $(48,64,16,32,4,4)$, $(32,64,16,32,4,4)$, $(32,48,16,32,4,4)$, $(48,80,16,32,4,4)$, $(48,64,24,48,4,4)$ \\
\hline

DeepONet
& All tasks
& \shortstack[l]{
$(64,32,3,64,3,\tanh)$, $(128,64,4,128,4,\tanh)$, $(256,64,4,128,4,\tanh)$, \\
$(128,128,4,128,4,\tanh)$, $(128,64,4,128,4,\mathrm{ReLU})$
} \\
\hline

UNet
& All tasks
& $(12)$, $(16)$, $(24)$, $(32)$, $(40)$ \\
\hline

\noalign{\vskip 0.8em}

Family & Task & Candidate vectors $(L,R,R_{\mathrm{neck}},d_e)$ \\
\hline

CNO
& \textit{Poisson}
& $(3,4,6,16)$, $(2,2,4,16)$, $(3,4,6,32)$, $(3,2,6,16)$, $(3,4,4,16)$ \\
& \textit{Wave}
& $(3,4,6,48)$, $(2,2,4,32)$, $(3,4,6,32)$, $(3,2,6,48)$, $(3,4,4,48)$ \\
& \textit{Allen--Cahn}
& $(3,4,8,48)$, $(2,2,6,32)$, $(3,4,8,32)$, $(3,2,8,48)$, $(3,4,4,48)$ \\
& \textit{Cont.\ translation}
& $(3,2,6,32)$, $(2,2,4,16)$, $(3,2,6,64)$, $(3,4,6,32)$, $(3,2,4,32)$ \\
& \textit{Disc.\ translation}
& $(3,5,4,32)$, $(2,2,4,16)$, $(3,5,4,64)$, $(3,2,4,32)$, $(3,5,8,32)$ \\
& \textit{Darcy}
& $(3,4,4,48)$, $(2,2,4,32)$, $(3,4,4,32)$, $(3,2,4,48)$, $(3,4,8,48)$ \\
& \textit{Navier--Stokes}
& $(3,1,8,32)$, $(2,1,6,16)$, $(3,1,8,64)$, $(3,4,8,32)$, $(3,1,4,32)$ \\
& \textit{Airfoil}
& $(4,1,8,48)$, $(3,1,6,32)$, $(4,1,8,32)$, $(4,4,8,48)$, $(4,1,4,48)$ \\
\hline
\end{tabular}%
}
\end{table*}

\paragraph{KNO candidate pool}

The KNO pool contains KNO-1, KNO-2, and KNO-3, defined in
Table~\ref{tab:kno-search-space}. The configurations share the feature width,
oscillator dimension, oscillator count, integration-step count $R$, canonical grid, and spectral
mode budget. They differ in layer count and allocation of
stages across layers.

\paragraph{Selected baseline architectures}
\label{app:selected-baseline-architectures}

Table~\ref{tab:baseline-selected-architectures} reports the architecture
selected for each architecture family and task by the stage-two criterion. Entries are hyperparameter vectors in the
notation of Table~\ref{tab:baseline-search-spaces} for baselines and Table~\ref{tab:kno-search-space} for KNO. AKOrN is omitted because
it uses one fixed architecture and is not subjected to architecture
selection.

\begin{table*}[!ht]
\centering
\caption{Architectures selected by the stage-two validation
criterion. Each body entry gives the hyperparameter vector.}
\label{tab:baseline-selected-architectures}
\scriptsize
\setlength{\tabcolsep}{3.0pt}
\renewcommand{\arraystretch}{1.15}

\begin{tabular}{@{}lccccc@{}}
\hline
Task
& FNO $(m,w,L,p)$
& RNO $(m,w)$
& ConvFNO $(m,w,c_{\mathrm U},c_{\mathrm{LSF}},
r_{\mathrm{lift}},r_{\mathrm{proj}})$
& DeepONet $(b,w_b,d_b,w_t,d_t,\sigma)$
& \\
\hline

\textit{Poisson}
& $(12,48,4,8)$
& $(12,64)$
& $(32,80,16,32,2,2)$
& $(64,32,3,64,3,\tanh)$
& \\

\textit{Wave}
& $(20,64,4,8)$
& $(24,128)$
& $(24,64,16,32,2,2)$
& $(128,64,4,128,4,\mathrm{ReLU})$
& \\

\textit{Allen--Cahn}
& $(20,96,5,0)$
& $(20,160)$
& $(48,80,16,32,4,2)$
& $(128,64,4,128,4,\mathrm{ReLU})$
& \\

\textit{Cont.\ translation}
& $(20,96,5,0)$
& $(20,128)$
& $(32,64,24,48,2,2)$
& $(128,64,4,128,4,\mathrm{ReLU})$
& \\

\textit{Disc.\ translation}
& $(20,96,5,0)$
& $(16,96)$
& $(32,64,24,48,2,2)$
& $(128,64,4,128,4,\mathrm{ReLU})$
& \\

\textit{Darcy}
& $(12,48,4,8)$
& $(20,160)$
& $(32,64,16,32,2,2)$
& $(128,64,4,128,4,\mathrm{ReLU})$
& \\

\textit{Navier--Stokes}
& $(20,96,5,0)$
& $(20,160)$
& $(48,80,16,32,4,4)$
& $(128,64,4,128,4,\mathrm{ReLU})$
& \\

\textit{Airfoil}
& $(16,48,4,12)$
& $(28,128)$
& $(32,64,16,32,4,4)$
& $(128,64,4,128,4,\mathrm{ReLU})$
& \\
\hline

\noalign{\vskip 0.8em}

Task
& UNet $(c)$
& CNO $(L,R,R_{\mathrm{neck}},d_e)$
& FNO-WT $(m,w,D,p)$
& FNO-DEQ $(m,w,D,p)$
& KNO \\
\hline

\textit{Poisson}
& $(32)$
& $(3,4,6,32)$
& $(16,24,3,8)$
& $(16,24,3,8)$
& KNO-2 \\

\textit{Wave}
& $(24)$
& $(3,4,6,32)$
& $(16,24,3,8)$
& $(16,24,3,8)$
& KNO-3 \\

\textit{Allen--Cahn}
& $(24)$
& $(3,4,8,48)$
& $(12,32,3,0)$
& $(12,32,3,0)$
& KNO-1 \\

\textit{Cont.\ translation}
& $(32)$
& $(3,4,6,32)$
& $(16,32,2,8)$
& $(16,32,2,8)$
& KNO-3 \\

\textit{Disc.\ translation}
& $(40)$
& $(3,5,4,64)$
& $(12,32,3,0)$
& $(16,24,3,8)$
& KNO-1 \\

\textit{Darcy}
& $(32)$
& $(3,4,8,48)$
& $(12,32,3,0)$
& $(8,40,3,8)$
& KNO-1 \\

\textit{Navier--Stokes}
& $(32)$
& $(3,1,8,32)$
& $(12,32,3,0)$
& $(12,32,3,0)$
& KNO-1 \\

\textit{Airfoil}
& $(12)$
& $(3,1,6,32)$
& $(8,40,3,12)$
& $(16,24,3,12)$
& KNO-1 \\
\hline
\end{tabular}
\end{table*}

\subsection{Optimization Protocol}
\label{app:optimization-protocol}

All KNO and baseline runs use AdamW with weight decay $10^{-5}$ and peak
learning rate $10^{-3}$. The learning rate increases linearly over the first
$20$ epochs and is then annealed to $10^{-6}$ using a cosine schedule.
Gradients are clipped to global norm $0.5$. Unless stated otherwise, the
batch size is $8$. Training and evaluation are performed in FP32.

The training objective and primary evaluation metric are both the
sample-averaged relative $L_2$ error in Eq.~\eqref{eq:relative-l2}. For every
run, the checkpoint attaining the lowest validation relative $L_2$ is
retained.

\subsection{Implementation Details}
\label{app:software-details}

KNO is implemented in PyTorch using components from the
\texttt{neuraloperator} library~\cite{kovachki2023neural}. FNO uses the
official \texttt{neuraloperator} implementation, RNO follows the authors'
implementation, DeepONet uses DeepXDE, and UNet follows the PDEBench
implementation. CNO follows the released \texttt{CNO2d\_classic}. ConvFNO follows
\cite{liu2025enhancing}. FNO-WT and FNO-DEQ use the same weight-tied FNO
block; FNO-WT applies $4$ explicit updates, whereas FNO-DEQ uses Anderson
iteration with phantom gradients.

Normalized spatial coordinates are supplied according to the standard
interface of each architecture. KNO appends normalized Cartesian coordinates
before lifting and uses circular padding in every spatial convolution.

\end{appendixpart}


\begin{thebibliography}{37}
\providecommand{\natexlab}[1]{#1}
\providecommand{\url}[1]{\texttt{#1}}
\expandafter\ifx\csname urlstyle\endcsname\relax
  \providecommand{\doi}[1]{doi: #1}\else
  \providecommand{\doi}{doi: \begingroup \urlstyle{rm}\Url}\fi

\bibitem[Anderson(1965)]{anderson1965iterative}
Donald~G Anderson.
\newblock Iterative procedures for nonlinear integral equations.
\newblock \emph{Journal of the ACM (JACM)}, 12\penalty0 (4):\penalty0 547--560, 1965.

\bibitem[Ashton et~al.(2024)Ashton, Mockett, Fuchs, Fliessbach, Hetmann, Knacke, Schonwald, Skaperdas, Fotiadis, Walle, et~al.]{ashton2024drivaerml}
Neil Ashton, Charles Mockett, Marian Fuchs, Louis Fliessbach, Hendrik Hetmann, Thilo Knacke, Norbert Schonwald, Vangelis Skaperdas, Grigoris Fotiadis, Astrid Walle, et~al.
\newblock Drivaerml: High-fidelity computational fluid dynamics dataset for road-car external aerodynamics.
\newblock \emph{arXiv preprint arXiv:2408.11969}, 2024.

\bibitem[Canuto et~al.(2006)Canuto, Hussaini, Quarteroni, and Zang]{canuto2006spectral}
Claudio Canuto, M~Youssuff Hussaini, Alfio Quarteroni, and Thomas~A Zang.
\newblock \emph{Spectral methods}, volume 285.
\newblock Springer, 2006.

\bibitem[Cao et~al.(2024)Cao, Goswami, and Karniadakis]{cao2024laplace}
Qianying Cao, Somdatta Goswami, and George~Em Karniadakis.
\newblock Laplace neural operator for solving differential equations.
\newblock \emph{Nature Machine Intelligence}, 6\penalty0 (6):\penalty0 631--640, 2024.

\bibitem[Chandra et~al.(2019)Chandra, Girvan, and Ott]{chandra2019continuous}
Sarthak Chandra, Michelle Girvan, and Edward Ott.
\newblock Continuous versus discontinuous transitions in the d-dimensional generalized kuramoto model: Odd d is different.
\newblock \emph{Physical Review X}, 9\penalty0 (1):\penalty0 011002, 2019.

\bibitem[Ciarlet(2002)]{ciarlet2002finite}
Philippe~G Ciarlet.
\newblock \emph{The finite element method for elliptic problems}.
\newblock SIAM, 2002.

\bibitem[Fresca et~al.(2021)Fresca, Dede’, and Manzoni]{fresca2021comprehensive}
Stefania Fresca, Luca Dede’, and Andrea Manzoni.
\newblock A comprehensive deep learning-based approach to reduced order modeling of nonlinear time-dependent parametrized pdes.
\newblock \emph{Journal of Scientific Computing}, 87\penalty0 (2):\penalty0 61, 2021.

\bibitem[Geng et~al.(2021)Geng, Zhang, Bai, Wang, and Lin]{geng2021training}
Zhengyang Geng, Xin-Yu Zhang, Shaojie Bai, Yisen Wang, and Zhouchen Lin.
\newblock On training implicit models.
\newblock \emph{Advances in neural information processing systems}, 34:\penalty0 24247--24260, 2021.

\bibitem[Gupta et~al.(2012)Gupta, Potters, and Ruffo]{gupta2012one}
Shamik Gupta, Max Potters, and Stefano Ruffo.
\newblock One-dimensional lattice of oscillators coupled through power-law interactions: Continuum limit and dynamics of spatial fourier modes.
\newblock \emph{Physical Review E—Statistical, Nonlinear, and Soft Matter Physics}, 85\penalty0 (6):\penalty0 066201, 2012.

\bibitem[Ja{\'c}imovi{\'c}(2024)]{jacimovic2024kuramoto}
Vladimir Ja{\'c}imovi{\'c}.
\newblock Kuramoto oscillators and swarms on manifolds for geometry informed machine learning.
\newblock \emph{Available at SSRN 4865439}, 2024.

\bibitem[Kendall and Gibbons(1962)]{kendall1962rank}
Maurice~George Kendall and Jean~Dickinson Gibbons.
\newblock Rank correlation methods.
\newblock 1962.

\bibitem[Kochkov et~al.(2020)Kochkov, Sanchez-Gonzalez, Smith, Pfaff, Battaglia, and Brenner]{kochkov2020learning}
Dmitrii Kochkov, Alvaro Sanchez-Gonzalez, Jamie~Alexander Smith, Tobias~Joachim Pfaff, Peter Battaglia, and Michael Brenner.
\newblock Learning latent field dynamics of pdes.
\newblock In \emph{Third Workshop on Machine Learning and the Physical Sciences (NeurIPS 2020)}, 2020.

\bibitem[Kovachki et~al.(2023)Kovachki, Li, Liu, Azizzadenesheli, Bhattacharya, Stuart, and Anandkumar]{kovachki2023neural}
Nikola Kovachki, Zongyi Li, Burigede Liu, Kamyar Azizzadenesheli, Kaushik Bhattacharya, Andrew Stuart, and Anima Anandkumar.
\newblock Neural operator: Learning maps between function spaces with applications to pdes.
\newblock \emph{Journal of Machine Learning Research}, 24\penalty0 (89):\penalty0 1--97, 2023.

\bibitem[Kuramoto(1984)]{kuramoto1984chemical}
Yoshiki Kuramoto.
\newblock Chemical turbulence.
\newblock In \emph{Chemical oscillations, waves, and turbulence}, pages 111--140. Springer, 1984.

\bibitem[LeVeque(2007)]{leveque2007finite}
Randall~J LeVeque.
\newblock \emph{Finite difference methods for ordinary and partial differential equations: steady-state and time-dependent problems}.
\newblock SIAM, 2007.

\bibitem[Li et~al.(2020)Li, Kovachki, Azizzadenesheli, Liu, Bhattacharya, Stuart, and Anandkumar]{li2020fourier}
Zongyi Li, Nikola Kovachki, Kamyar Azizzadenesheli, Burigede Liu, Kaushik Bhattacharya, Andrew Stuart, and Anima Anandkumar.
\newblock Fourier neural operator for parametric partial differential equations.
\newblock \emph{arXiv preprint arXiv:2010.08895}, 2020.

\bibitem[Li et~al.(2023{\natexlab{a}})Li, Huang, Liu, and Anandkumar]{li2023fourier}
Zongyi Li, Daniel~Zhengyu Huang, Burigede Liu, and Anima Anandkumar.
\newblock Fourier neural operator with learned deformations for pdes on general geometries.
\newblock \emph{Journal of Machine Learning Research}, 24\penalty0 (388):\penalty0 1--26, 2023{\natexlab{a}}.

\bibitem[Li et~al.(2023{\natexlab{b}})Li, Kovachki, Choy, Li, Kossaifi, Otta, Nabian, Stadler, Hundt, Azizzadenesheli, et~al.]{li2023geometry}
Zongyi Li, Nikola Kovachki, Chris Choy, Boyi Li, Jean Kossaifi, Shourya Otta, Mohammad~Amin Nabian, Maximilian Stadler, Christian Hundt, Kamyar Azizzadenesheli, et~al.
\newblock Geometry-informed neural operator for large-scale 3d pdes.
\newblock \emph{Advances in Neural Information Processing Systems}, 36:\penalty0 35836--35854, 2023{\natexlab{b}}.

\bibitem[Lipton et~al.(2021)Lipton, Mirollo, and Strogatz]{lipton2021kuramoto}
Max Lipton, Renato Mirollo, and Steven~H Strogatz.
\newblock The kuramoto model on a sphere: Explaining its low-dimensional dynamics with group theory and hyperbolic geometry.
\newblock \emph{Chaos: An Interdisciplinary Journal of Nonlinear Science}, 31\penalty0 (9), 2021.

\bibitem[Liu et~al.(2025)Liu, Murari, Liu, Li, Budd, and Sch{\"o}nlieb]{liu2025enhancing}
Chaoyu Liu, Davide Murari, Lihao Liu, Yangming Li, Chris Budd, and Carola-Bibiane Sch{\"o}nlieb.
\newblock Enhancing fourier neural operators with local spatial features.
\newblock \emph{arXiv preprint arXiv:2503.17797}, 2025.

\bibitem[Liu et~al.(2026)Liu, Yang, and Chen]{yangriesz}
Shouyi Liu, Xiaokang Yang, and Yuntian Chen.
\newblock Riesz neural operator for solving partial differential equations.
\newblock In \emph{The Fourteenth International Conference on Learning Representations}, 2026.
\newblock URL \url{https://openreview.net/forum?id=Vjw7q1quNt}.

\bibitem[Lohe(2009)]{lohe2009non}
MA2539317 Lohe.
\newblock Non-abelian kuramoto models and synchronization.
\newblock \emph{Journal of Physics A: Mathematical and Theoretical}, 42\penalty0 (39):\penalty0 395101, 2009.

\bibitem[Lu et~al.(2021)Lu, Jin, Pang, Zhang, and Karniadakis]{lu2021learning}
Lu~Lu, Pengzhan Jin, Guofei Pang, Zhongqiang Zhang, and George~Em Karniadakis.
\newblock Learning nonlinear operators via deeponet based on the universal approximation theorem of operators.
\newblock \emph{Nature machine intelligence}, 3\penalty0 (3):\penalty0 218--229, 2021.

\bibitem[Lu et~al.(2026)Lu, Chen, Xu, Li, Zhu, and Zheng]{lu2026solving}
Wenbin Lu, Yihan Chen, Junnan Xu, Wei Li, Junwei Zhu, and Jianwei Zheng.
\newblock Solving partial differential equations via radon neural operator.
\newblock \emph{Advances in Neural Information Processing Systems}, 38:\penalty0 157027--157067, 2026.

\bibitem[Markdahl et~al.(2021)Markdahl, Proverbio, Mi, and Goncalves]{markdahl2021almost}
Johan Markdahl, Daniele Proverbio, La~Mi, and Jorge Goncalves.
\newblock Almost global convergence to practical synchronization in the generalized kuramoto model on networks over the n-sphere.
\newblock \emph{Communications Physics}, 4\penalty0 (1):\penalty0 187, 2021.

\bibitem[Marwah et~al.(2023)Marwah, Pokle, Kolter, Lipton, Lu, and Risteski]{marwah2023deep}
Tanya Marwah, Ashwini Pokle, J~Zico Kolter, Zachary Lipton, Jianfeng Lu, and Andrej Risteski.
\newblock Deep equilibrium based neural operators for steady-state pdes.
\newblock \emph{Advances in Neural Information Processing Systems}, 36:\penalty0 15716--15737, 2023.

\bibitem[Medvedev(2018)]{medvedev2018continuum}
Georgi~S Medvedev.
\newblock The continuum limit of the kuramoto model on sparse random graphs.
\newblock \emph{arXiv preprint arXiv:1802.03787}, 2018.

\bibitem[Miyato et~al.(2025)Miyato, L{\"o}we, Geiger, and Welling]{miyato2025artificial}
Takeru Miyato, Sindy L{\"o}we, Andreas Geiger, and Max Welling.
\newblock Artificial kuramoto oscillatory neurons.
\newblock In \emph{International Conference on Learning Representations}, volume 2025, pages 44278--44322, 2025.

\bibitem[Moukalled et~al.(2015)Moukalled, Mangani, and Darwish]{moukalled2015finite}
Fadl Moukalled, Luca Mangani, and Marwan Darwish.
\newblock The finite volume method.
\newblock In \emph{The finite volume method in computational fluid dynamics: An advanced introduction with OpenFOAM{\textregistered} and Matlab}, pages 103--135. Springer, 2015.

\bibitem[Raonic et~al.(2023)Raonic, Molinaro, De~Ryck, Rohner, Bartolucci, Alaifari, Mishra, and De~B{\'e}zenac]{raonic2023convolutional}
Bogdan Raonic, Roberto Molinaro, Tim De~Ryck, Tobias Rohner, Francesca Bartolucci, Rima Alaifari, Siddhartha Mishra, and Emmanuel De~B{\'e}zenac.
\newblock Convolutional neural operators for robust and accurate learning of pdes.
\newblock \emph{Advances in Neural Information Processing Systems}, 36:\penalty0 77187--77200, 2023.

\bibitem[Ronneberger et~al.(2015)Ronneberger, Fischer, and Brox]{ronneberger2015u}
Olaf Ronneberger, Philipp Fischer, and Thomas Brox.
\newblock U-net: Convolutional networks for biomedical image segmentation.
\newblock In \emph{International Conference on Medical image computing and computer-assisted intervention}, pages 234--241. Springer, 2015.

\bibitem[Sacchetti(2020)]{sacchetti2020derivation}
Andrea Sacchetti.
\newblock Derivation of the tight-binding approximation for time-dependent nonlinear schr{\"o}dinger equations.
\newblock In \emph{Annales Henri Poincar{\'e}}, volume~21, pages 627--648. Springer, 2020.

\bibitem[Song et~al.(2026)Song, Keller, Brodjian, Miyato, Yue, Perona, and Welling]{song2026kuramoto}
Yue Song, Andy Keller, Sevan Brodjian, Takeru Miyato, Yisong Yue, Pietro Perona, and Max Welling.
\newblock Kuramoto orientation diffusion models.
\newblock \emph{Advances in Neural Information Processing Systems}, 38:\penalty0 4047--4076, 2026.

\bibitem[Torre(2014)]{torre201405}
Charles~G Torre.
\newblock 05 the continuum limit and the wave equation.
\newblock \emph{Continuum}, 1:\penalty0 1--2004, 2014.

\bibitem[Wen et~al.(2022)Wen, Li, Azizzadenesheli, Anandkumar, and Benson]{wen2022u}
Gege Wen, Zongyi Li, Kamyar Azizzadenesheli, Anima Anandkumar, and Sally~M Benson.
\newblock U-fno—an enhanced fourier neural operator-based deep-learning model for multiphase flow.
\newblock \emph{Advances in Water Resources}, 163:\penalty0 104180, 2022.

\bibitem[Wu et~al.(2026)Wu, Guo, Li, Dou, Long, He, and Matusik]{wu2026geopt}
Haixu Wu, Minghao Guo, Zongyi Li, Zhiyang Dou, Mingsheng Long, Kaiming He, and Wojciech Matusik.
\newblock Geopt: Scaling physics simulation via lifted geometric pre-training.
\newblock \emph{arXiv preprint arXiv:2602.20399}, 2026.

\bibitem[Xiao et~al.(2026)Xiao, Wang, Han, Shan, and Li]{xiao2026kuramoto}
Mingqing Xiao, Yansen Wang, Dongqi Han, Caihua Shan, and Dongsheng Li.
\newblock Kuramoto oscillatory phase encoding: Neuro-inspired synchronization for improved learning efficiency.
\newblock \emph{arXiv preprint arXiv:2604.07904}, 2026.

\end{thebibliography}
\end{document}